# Electron Photodetachment from Aqueous Anions.

# II. Ionic Strength Effect on Geminate Recombination Dynamics and Quantum Yield for Hydrated Electron. [1]

Myran C. Sauer, Jr., [*] Ilya A. Shkrob, [*] Rui Lian,

Robert A. Crowell, David M. Bartels, [a)]

*Chemistry Division , Argonne National Laboratory, Argonne, IL 60439*

Stephen E. Bradforth

*Department of Chemistry, University of Southern California, Los Angeles, CA 90089*

**Abstract**

In concentrated solutions of $NaClO_4$ and $Na_2SO_4$, the quantum yield for free electron generated by detachment from photoexcited anions (such as $I^-$, $OH^-$, $ClO_4^-$, and $SO_3^{2-}$) linearly decreases by 6-12% per 1 M ionic strength. In 9 M sodium perchlorate solution, this quantum yield decreases by roughly an order of magnitude. Ultrafast kinetic studies of 200 nm photon induced electron detachment from $Br^-$, $HO^-$ and $SO_3^{2-}$ suggest that the prompt yield of thermalized electron does not change in these solutions; rather, the ionic strength effect originates in more efficient recombination of geminate pairs. Within the framework of the recently proposed mean force potential (MFP) model of charge separation dynamics in such photosystems, the observed changes are interpreted as an increase in the short-range attractive potential between the geminate partners. Association of sodium cation(s) with the electron and the parent anion is suggested as the most likely cause for the observed modification of the MFP. Electron thermalization kinetics suggest that the cation associated with the parent anion (by ion pairing and/or ionic atmosphere interaction) is passed to the detached electron in the course of the photoreaction. The precise atomic-level mechanism for the ionic strength effect is presently unclear; any further advance is likely to require the development of an adequate quantum molecular dynamics model.

\*   To whom correspondence should be addressed: *Tel* 630-252-9516, *FAX* 630-2524993, *e-mail:* shkrob@anl.gov.
a)  present address: Radiation Laboratory, University of Notre Dame, Notre Dame, Indiana 46556, *e-mail:* bartels.5@nd.edu.

## 1. Introduction.

In Part I of this series, [1] estimates for absolute quantum yields (QYs) and cross sections for electron photodetachment from miscellaneous aqueous anions were given. With a single exception of perchlorate, these measurements were carried out in dilute (< 50 mM) solutions of these CTTS (charge transfer to solvent) active anions. In this work, we explore the electron dynamics for photoexcited anions in high ionic strength solutions.

The motivation for such a study is provided by the recent ultrafast kinetic studies of electron photodetachment from halide [2,3,4] and pseudohalide [4,5,6] anions. These studies suggest the existence of a weak attractive mean force potential (MFP) between the residual radical (such as HO or a halogen atom) and the ejected electron that localizes in its vicinity. This attractive potential causes bimodality of the electron decay kinetics: in the first 10-50 ps these kinetics are exponential (due to the fast escape and recombination of the electrons situated near the bottom of the potential well); [7] these rapid kinetics are succeeded by a slower $t^{-1/2}$ decay due to the diffusional escape and recombination of hydrated electrons which are thermally emitted from this potential well. [2,5,7] For polyvalent anions, only these slow kinetics are observed; [4,8] the fast exponential component is lacking. This is not surprising since long-range Coulomb repulsion between the electron and a radical anion (derived from the parent polyvalent anion) should be much stronger than short-range, weak attraction between the electron and a neutral radical (derived from the parent monovalent anion). For the latter type of geminate pairs, the MFP is thought to originate mainly through the polarization of the residue (e.g., halide atom) by the electron, [9,10] though electron - dipole interaction may also be significant for pairs generated by electron photodetachment from polyatomic anions. According to Bradforth and coworkers, [9] for geminate pairs derived from monovalent anions, the strength of the interaction (as estimated from MFP model kinetic fits) increases with the polarizability of the radical/atom, being maximum for iodine atoms. For chloride, the MFP has been obtained theoretically, using quantum molecular dynamics simulations and umbrella sampling. [10] Unfortunately, these simulations do not point to any *specific* interactions that are responsible for the MFP: the latter emerges as a sum total of many interparticle interactions. Still, it is reassuring that the strength of this simulated MFP (which is a few *kT* units) is comparable to the strength of the MFP that was extracted by Bradforth and coworkers from the experimental kinetics for aqueous iodide. [2]

These MFP models were found to quantitatively account for (i) the temporal profile of the geminate recombination kinetics, [2,4,5] (ii) the changes observed in these kinetics with increasing photoexcitation energy (direct ionization that occurs at higher energies broadens the spatial distribution of photoejected electrons), [11] (iii) temperature effects on the decay kinetics of electrons generated in 200 nm photoexcitation of hydroxide. [5] The MFP model also qualitatively accounts for the trends in the kinetics observed for homologous anions (e.g., the correlation between the potential well depth and the polarizability of the radical), [9] as well as some trends observed for nonaqueous and mixed solvents. [3a,b]

Impressive as this record may appear, there is no direct, unambiguous way in which the MFP profile can be extracted from the experimental data, and this leaves room for fudging the MFP parameters and the initial electron distribution to accommodate any of the trends observed. E. g., to "explain" temperature effects on the geminate kinetics for $\left(OH, e_{aq}^-\right)$ pairs generated in electron photodetachment from hydroxide it was postulated that the MFP for this pair becomes shallower and more diffuse with the increasing temperature. [5] While the entire set of 8 °C to 90 °C kinetics can be consistently fit with these assumptions, there is currently no way of demonstrating experimentally that the MFP behaves as postulated or that this attractive potential even exists. This uncertainty pertains to other studies: in the end, it is not yet clear whether the MFP model is a correct (though simplified) picture that captures peculiarities of electron dynamics for photoexcited CTTS anions or merely a recipe for producing kinetic profiles that resemble those observed experimentally.

We believe that the MFP model does capture the reality, albeit incompletely. E.g., the existence of a short-range interaction that temporarily stabilizes close pairs is suggested by experimental [12] and theoretical studies [13] of Na$^-$ CTTS by Schwartz and coworkers. These authors showed that electrons in the close pairs (observed on the short time scale) responded differently to IR photoexcitation than the electrons in more distant pairs. The short-term electron dynamics observed in the sodide photosystem had been viewed as the dissociation of these close pairs (in which the electron and sodium atom reside in a single solvent cavity) to solvent separated pairs (in which the species are separated by 1-2 solvent molecules) that subsequently decay by diffusional migration of the partners to the bulk.

The purpose of the present study was to validate the use of the MFP model for aqueous anions. The approach was to modify this potential in a predictable way, through ionic atmosphere screening of electrostatic charges. It was expected that for geminate pairs derived from monovalent anions, the screening of the electron would weaken the electrostatic interactions of this electron with its neutral partner and the survival probability $\Omega_\infty$ of electrons that escape geminate recombination would thereby increase. Conversely, for polyvalent anions, this probability would decrease because the ionic atmosphere would screen the Coulomb repulsion between the electron and its negatively charged partner, the radical anion. As demonstrated below, these expectations were not realized.

The experimental results show that both the survival probability and the quantum yield of free electron *decrease* with the ionic strength for all of the systems studied, whereas the prompt QY for the solvated electrons does not change. The magnitude of this decrease is the same for monovalent and divalent anions and changes little from one photosystem to another; it also varies little with the salt that is used to change the ionic strength. While similar magnitude of the decrease in the survival probability $\Omega_\infty$ of the electron with ionic strength was attained for all photosystems, the rate for the attainment of this decrease varied considerably between different photosystems.

Though we cannot presently suggest a detailed model that accounts for these observations in their totality, it is certain that ion screening cannot account for these results even at the qualitative level. We suggest that the effect originates through the modification of the MFP via alkali cation association with the solvated electron and the parent anion (similar to the case of ferrocyanide CTTS considered by Bradforth and coworkers); [8] somehow, this association makes the short-range attraction stronger, for all anions.

In the view of this latter conclusion, most of this paper constitutes an extended proof of the assertions given above. That task required combining several spectroscopic techniques, including ultrafast pump-probe spectroscopy, laser flash photolysis, and pulse radiolysis. To reduce the length of the manuscript, Figs. 1S to 9S are placed in the Supporting Information.

**2. Experimental.**

*Nanosecond flash photolysis and QY measurement.* The setup used to determine the QY for free electron formation has been described in Part I of this series. [1] Fifteen nanosecond fwhm, 1-20 mJ pulses from an ArF (193 nm) or KrF (248 nm) excimer laser (Lamda Physik LPX 120i) were used to photolyze $N_2$- or $CO_2$- saturated aqueous solutions. The laser and analyzing light beams were crossed at an angle of $30^o$ inside a 1.36 mm optical path cell. A fast silicon photodiode equipped with a video amplifier terminated into a digital signal analyzer were used to sample the transient absorbance kinetics (> 3 ns). Two pyroelectric energy meters were used to measure the power of the incident and transmitted UV light during the kinetic sampling. 200-500 ml of the solution being studied was circulated through the cell using a peristaltic pump. The typical flow rate was 2-3 mL/min; the repetition rate of the laser was 1.7 Hz. Purified water with conductivity < 2 nS/cm was used to prepare the aqueous solutions. The UV spectra were obtained using a dual beam spectrophotometer (OLIS/Cary 14).

The typical QY measurement included determination of laser transmission $T$ through the sample, transient absorbance $\Delta OD_\lambda$ of the photoproduct at wavelength $\lambda$ of the analyzing light, the laser energy $I_{abs}$ absorbed by the sample, and the incident beam energy $I_0$. The QY was determined from the initial slope of $\Delta OD_\lambda$ vs. $I_{abs}$ that was corrected for the window transparency and noncolinear beam geometry. If not stated otherwise, the absorbance of hydrated electron at $\lambda$=700 nm ($\varepsilon_{700}$=20560 M$^{-1}$ s$^{-1}$ [14,15]) at the end of the UV pulse ($t$=30 ns) was used to estimate the electron yield. The anion concentration and light fluence were chosen so that the decay half time of this electron (due to cross recombination) was > 5 µs.

*Ultrafast laser spectroscopy.* The pico- and femto- second kinetic measurements reported below were obtained using a 1 kHz Ti:sapphire setup at the ANL whose details are given in refs. [4,5,6]. Stretched 2 nJ pulses from a Ti:sapphire oscillator were amplified to 4 mJ in a home-built two-stage multipass Ti:sapphire amplifier. The amplified pulses were passed through a grating compressor that yielded Gaussian pulses of 60 fs FWHM and 3 mJ centered at 800 nm. The amplified beam was split into three parts of approximately equal energy. One beam was used to generate 800 nm probe pulses while the other two were used to generate the 200 nm (fourth harmonic) pump pulses by upconversion of the third harmonic. Up to 20 µJ of the 200 nm light was produced this way (300-350 fs FWHM pulse). The pump power, before and after the sample, was determined using a calibrated thermopile power meter (Ophir Optronics model 2A-SH). The probe pulse was delayed in time as much as 600 ps using a 3-pass 25 mm delay line. The pump and probe beams were perpendicularly polarized, focused to round spots of 50 µm and 200-300 µm in radius, respectively, and overlapped in the sample at 6.5$^{o}$. Commonly in the pump-probe experiments, the pump beam envelopes the probe beam. We reversed this geometry because of the poor quality of the 200 nm beam ("hot spots" in the beam profile). This reverse geometry minimizes the sensitivity of the absorption signal to fluctuations of the "hot spot" pattern and the walk-off of the probe beam relative to the pump beam (which was typically 5-10 µm). The trade-off was a considerable reduction in the signal magnitude that was more than compensated by improved signal-to-noise ratio and shot-to-shot stability.

The probe and the reference signals were detected with fast silicon photodiodes, amplified, and sampled using home-built electronics. The typical standard deviation for a pump-probe measurement was $10^{-5}$ *ΔOD*, and the "noise" was dominated by the variation of the pump intensity and flow instabilities in the jet. The vertical bars shown in the kinetics (e.g., Fig. 1) represent 95% confidence limits for each data point. Typically, 150-200 points acquired on a quasi-logarithmic grid were used to obtain these kinetics. The experiments were carried out in flow, using a 150 μm thick high-speed jet. An all 316 stainless steel and Teflon flow system was used to pump the solution.

*Pulse radiolysis - transient absorption spectroscopy.* 20 MeV, 30 ps FWHM electron beam pulses of 20 nC from the Argonne LINAC were used to radiolyze $N_2$-saturated aqueous solutions being studied in a 20 mm optical path cell (the setup used in this work was similar to that described in ref. [16]). The analyzing light from a pulsed Xe arc lamp was coaxial with the electron beam and traveled in the opposite direction. A set of 10 nm fwhm band pass interference filters was used for wavelength selection. The kinetic traces were sampled and averaged at a repetition rate of 0.3 Hz. The typical time sampling interval was 0.2 ns per point. Cerenkov light signal was subtracted from the kinetic traces. A Faraday cup was used to monitor pulse-to-pulse stability of the electron beam. While the accumulation of radiolytic products slowly decreased the electron life time during the acquisition of the absorption spectra, this shortening had negligible effect on the absorbance at $t=5$ ns, the delay time at which these spectra were obtained.

*Materials and solution properties.* Reagents of the highest purity available from Aldrich were used without further purification. The solutions were made of crystalline salts with exception of hydroxide solutions that were made using high-purity 0.989 N analytical standard KOH.

Density, viscosity, activity and ion diffusivity charts for sodium sulfate and sodium perchlorate solutions were taken from the literature (some of the relevant data are given in Figs. 1S, 2S, and 3S). [17,21] In saturated room temperature solution, these molarities and the densities are 1.84 M and 1.21 g/ml for sodium sulfate and 9.32 M and 1.68 g/ml for sodium perchlorate, respectively. [18] The refractive indices of these solutions change very little relative to that of neat water (at saturation, 1.368 for $NaClO_4$ [19] and 1.364 for $Na_2SO_4$ [20] vs. 1.333 for water; [20] for the sodium line). The density of the solution scales approximately linearly with the molarity of the salt (Figs. 1S(a) and 3S(a)). [17] This increase in the density results in a higher stopping power for electrons in briny solutions and has important implications for pulse radiolytic experiments. The quantity that changes most with the salinity is the bulk viscosity: from 1 cP for water to 1.9 cP for sodium sulfate (Fig. 1S(a)) and to 8 cP for sodium perchlorate (Fig. 3S(a)), for saturated solutions (for the latter salt, most of the increase in the viscosity occurs between 6 and 9 M, Fig. 3S(a)). Microscopic diffusion coefficients for the corresponding ions change less than these bulk viscosities. [21] In 2.5 M $NaClO_4$ and 1.8 M $Na_2SO_4$, the sodium cation migrates ca. 30% more slowly than in water (Figs. 1S(b) and 3S(b)). In 0.5-2 M $Na_2SO_4$, the diffusion coefficient of sulfate is constant, also ca. 30 % lower than in water (Fig. 1S(b)). Thus, the drag of the ionic atmosphere on the migration of these ions is relatively weak. The ionic strength given below was calculated assuming the absence of ion pairing. For sodium sulfate, Raman spectroscopy [22], dielectric permittivity [23], ultrasonic absorption [24], and conductivity [25] studies suggest the occurrence of noncontact ion pairs in which the sulfate anion and the sodium cation are separated by one or two water molecules. Since the ion activities rapidly decrease with the salt concentration (Fig. 2S(a)) [23], the association constant decreases by two orders of magnitude from dilute to saturated solution (see Fig. 3 in ref. [23]), and ion pair formation peaks at ca. 0.2 M (Fig. 2S(b)) [23]. When this ion pairing is included, the ionic strength of the $Na_2SO_4$ solutions is ca. 2.83 times the molarity of the sodium sulfate (Fig. 2S(b)).

**3. Results.**

This section is organized as follows. First, the choice of (inert) salts used to change the ionic strength of CTTS anion solutions under study is justified. It is shown that only two types of such salts, perchlorates and sulfates, can be used for studies of the ionic strength effect in the CTTS photosystems (section 3.1). The addition of these salts shifts the absorption spectra of CTTS anions and the hydrated electron to the blue (sections 3.2 and 3.3, respectively). Since the molar absorptivities of the parent anion and the electron are therefore concentration dependent, this spectral shift has to be taken into the account for our QY measurements. Pulse radiolysis - transient absorbance spectroscopy was used to observe the effect of ionic strength on the shape of the electron spectrum in the visible (section 3.3); this approach, however, did not give the absolute molar absorptivity for the electron. To forgo this difficulty, first the product $\phi\varepsilon_{700}$ of the QY and the molar absorptivity of the electron was obtained for several CTTS systems as a function of the ionic strength of the photolyzed solution (section 3.4). Then, the iodide photosystem (248 nm excitation) was used to obtain the electron and iodine atom yields separately; these two sets of measurements were used to estimate the ionic strength dependence of the electron absorptivity at the observation wavelength of 700 nm (section 3.5). The latter data, taken together with pulse radiolysis data of section 3.2, suggest that the oscillator strength of the electron transition in the visible does not change with the ionic strength. Using the estimates for $\varepsilon_{700}$ obtained in the iodide experiment, absolute quantum yields $\phi$ were obtained as a function of the ionic strength $I$ (section 3.5). All of these dependencies were linear, with a negative slope of 6-10% per 1 M of ionic strength. Ultrafast pump-probe spectroscopy was used to demonstrate that this effect originates through a perturbation in the recombination dynamics of the corresponding geminate pairs; there is no ionic strength effect on the *prompt* yield of photoelectrons in the CTTS excitation (section 3.6). The kinetics for electron photodetachment from bromide are analyzed using an MFP model of geminate recombination based on the work of Shushin [7]. These analyses suggest that the short-range attractive potential between the geminate partners increases with ionic strength.

### *3.1. The choice of the ion atmosphere.*

In our experiments, the ionic strength of aqueous solutions was changed by addition of sodium sulfate and sodium perchlorate. This choice was dictated by the following practical considerations:

First, most anions are reactive towards the radical generated in the course of electron photodetachment. Typical reactions include addition (e.g., halogen atoms and OH radicals readily add to halide anions) and oxidation (OH, $SO_3^-$ and $SO_4^-$ are very strong oxidizers). The sulfate and perchlorate are two isoelectronic anions for which no addition or oxidation reactions have ever been reported. This guarantees that the salts used for changing the ionic strength are not *chemically* involved in the electron detachment and/or geminate recombination.

Second, the anions must be sufficiently weak absorbers of photoexcitation light, even at molar concentration of these anions, so that most of the photoexcitation light is absorbed by the CTTS anions of interest which are present in a relatively low concentration. The latter consideration forced us to use CTTS anions that were good light absorbers ($\varepsilon > 500$ M$^{-1}$ cm$^{-1}$) at the photoexcitation wavelength, such as iodide at 248 nm and hydroxide at 193 nm (see Table 1 in ref. [1]) and to use sufficiently high concentrations of these anions so that < 0.1% of the light was absorbed by the buffer salt (in flash photolysis experiments). Nanosecond observations suggested that neither sulfate nor perchlorate yield electrons by single 248 nm photon excitation, even in saturated solutions (the QY is < 10$^{-4}$). For 193 nm photoexcitation, sulfate is a relatively strong light absorber ($\varepsilon_{193} = 46\pm7$ M$^{-1}$ cm$^{-1}$ [1]), and only perchlorate can be used ($\varepsilon_{193} = 0.565\pm0.007$ M$^{-1}$ cm$^{-1}$ [1]). Furthermore, as shown in Part I of this series, [1] perchlorate yields electrons by one- or two- 193 nm photon excitation (the one-photon QY is 4x10$^{-3}$ [1]). Since the absorption band of perchlorate shifts to the blue with the perchlorate concentration, at least 20-50% of the 193 nm light is observed by 1-2 mM hydroxide, even in 9 M NaClO$_4$ solution. In dilute aqueous solution, the QY for hydroxide is ca. 0.11, [1,26,27] and < 3% of the electron absorbance signal was from perchlorate, for all concentrations of the latter. Still, corrections were necessary to determine QYs for the OH$^-$/NaClO$_4$ photosystem, as explained in section 3.4.

In 200 nm short-pulse photoexcitation of 0.74 M sulfate, a transient from hydrated electron was observed at 800 nm (Fig. 1(a), trace (i)). The transmission of the 200 nm light through the 150 μm jet was 0.74, which gives an estimate of 12 M$^{-1}$ cm$^{-1}$ for the molar absorptivity of sulfate at 200 nm. This value is in the correct range for this anion, judging from our 193 nm estimate of the molar absorptivity (46±7 M$^{-1}$ cm$^{-1}$ [1]) and the spectra given in ref. [28] This absorption from the electron scaled linearly with the photon fluence of 200 nm light (Fig. 4S(a)), and the observed photoelectron decay kinetics, - which did not depend on the laser power (Fig. 4S(b)), - resembled the kinetics observed for other polyvalent anions (such as SO$_3^-$ [4]). From the plot given in Fig. 4S(a), a relatively large prompt quantum yield of 0.406 was obtained for the electron at 10 ps (that decreased to 0.27 at 500 ps). A free electron QY of 0.83 was obtained for sulfate at 193 nm, [1,26,27] and we believe that the kinetics shown in Fig. 1(a) are from photoexcited sulfate, rather than an impurity. To make sure that sulfite (that has large absorptivity at 200 nm) is not present in the solution, we oxidized it by prolonged bubbling of the sulfate solution with oxygen; no change was observed in the electron absorbance. Other impurities (<0.1 wt%) specified by the manufacturer (such as chloride and metal cations) are weaker absorbers of 200 nm light than sulfate.

A 9.25 M perchlorate solution also yielded 800 nm absorbance from the electron by 200 nm photoexcitation (Fig. 1(b)). The transmission of this solution at 200 nm was 0.67 (optical density of 0.18) which is equivalent to molar absorptivity of 1.3 $M^{-1}$ $cm^{-1}$ vs. 0.26 $M^{-1}$ $cm^{-1}$ obtained for 193 nm. [1] The transient absorption signal is either from biphotonic excitation of perchlorate (as suggested by our 193 nm results [1]) or one-photon excitation of an impurity in the perchlorate; the signal was too small to distinguish between these two possibilities. The involvement of impurity is likely: e.g., decay kinetics shown in Fig. 1(a) look remarkably similar to the decay kinetics observed for 5.4 M chloride (Fig. 1(a), solid line).

Since the optical density of 30 mM bromide ( $\varepsilon_{200}$ of $10^4$ $M^{-1}$ $cm^{-1}$) across the 150 μm jet was 4.5, the absorption of light by perchlorate was < 4 % of that absorbed by bromide. Given that the prompt QY for electron in 200 nm photoexcitation of bromide is ca. 0.91, [4] the contribution of perchlorate to the electron generation in 30 mM bromide was negligible. In fact, no change in the kinetic profiles was observed when the bromide concentration was varied between 20 and 160 mM (see Fig. 7S in section 3.6.1).

*3.2. The effect of salts on the CTTS bands of anions.*

It is known from previous studies that addition of both ionic and nonionic solutes to water shifts the CTTS absorption bands of anions to the blue. According to Blandamer and Fox [29] and Stein and Treinin, [30] this shift changes neither the shape of the absorption band nor the molar absorptivity of the anion at the band maximum. The latter authors studied the effect of $NaClO_4$ and $Na_2SO_4$ on the absorption of aqueous iodide; however, only band maxima positions were reported. To obtain more detailed data, UV spectra of 0.7 mM NaI in 1 mm and 10 mm optical path cells were collected as a function of salt concentration. For λ>205-210 nm, the (low energy sub-) band was perfectly Gaussian; the band width (314±4 meV) and the molar absorptivity at the maximum ($1.1 \times 10^4$ $M^{-1}$ $cm^{-1}$) changed < 2% with the salt concentration (Fig. 2(a)). For both salts, the blue shift *(-Δλ)* linearly increased with the nominal ionic strength, and the band shifts given by Stein and Treinin [30] were in perfect agreement with those obtained in this work (Fig. 2(b)). In energy units, the band shifts for sodium perchlorate and sodium sulfate are 26 meV and 13.4 meV per 1 M ionic strength. Thus, in 9 M $NaClO_4$ solution, the band shift is ca. 10 nm. Such a large shift strongly changes the molar absorptivity of the anion at the low energy band tail. Fig. 3(a) shows the plot of the molar extinction coefficient $\varepsilon_{248}$ of iodide at 248 nm as a function of the ionic strength. These data were obtained from spectrophotometric data of Fig. 2(a) and compared with values estimated from the transmission of 248 nm light from the KrF laser. These two methods gave similar results: As seen from this plot, an increase in the ionic strength causes a large decrease in the molar absorptivity. For 9 M $NaClO_4$ solution, the molar absorptivity of iodide decreases by more than a decade. Only for this $I^-/NaClO_4$ photosystem was the blue shift sufficiently large to produce nonlinear dependence of the molar absorptivity (at the band edge) on the salt concentration. For other photosystems studied at 248 nm and 193 nm, the molar absorptivity linearly decreased with the salinity. For hydroxide at 193 nm (where the anion has a band maximum), the molar absorptivity changed from 3000 $M^{-1}$ $cm^{-1}$ in water to 900 $M^{-1}$ $cm^{-1}$ in 9 M $NaClO_4$ solution (see Fig. 3(b)). For sulfite at 248 nm, the molar absorptivity changed from 40 $M^{-1}$ $cm^{-1}$ in water to 25 $M^{-1}$ $cm^{-1}$ in 1.8 M $Na_2SO_4$ (see Fig. 3(b)). Thus, for all photosystems (including 200 nm photosystems) the absorptivity of the solution at the excitation wavelength systematically decreases with the ionic strength.

### 3.3. The effect of salts on the electron absorption.

It is well known that hydrated electrons and halide anions show similar trends in their absorption spectra as a function of solvent polarity, temperature, etc. [1,31] Thus, it is not surprising that the absorption spectrum of the hydrated electron shifts to the blue in concentrated salt solutions. For 10 M $NaClO_4$, a large shift of the band maximum, from 710 nm to 605 nm, was reported by Anbar and Hart. [32,33] Since the free electron concentration in our time-resolved QY measurement was determined from the hydrated electron absorbance at 700 nm, one needs to know how the extinction coefficient of this hydrated electron changes with the ionic strength of the solution.

To that end, we obtained the spectra of hydrated electrons generated by pulse radiolysis of sodium sulfate (Fig. 4(a)) and sodium perchlorate (Fig. 4(b)) solutions for different concentrations of these salts (same dose for all systems). These spectra were obtained 5-10 ns after a 30 ps pulse of 20 MeV electrons. In water, the absorption spectrum $S(E)$ of the electron as a function of photon energy $E$ can be approximated by a Lorentzian-Gaussian dependence: $S(E) = S_{max} \exp\left(-\left[(E-E_{max})/\Gamma_G\right]^2\right)$ for $E < E_{max}$ and $S(E) = S_{max}/\left(1+\left[(E-E_{max})/\Gamma_L\right]^2\right)$ for $E > E_{max}$, where $S_{max}$ and $E_{max}$ are the signal magnitude and photon energy at the band maximum, respectively, and $\Gamma_G$ and $\Gamma_L$ are the Gaussian and Lorentzian widths, respectively. [15,33] For light water, our recent measurement (which agrees reasonably well with the previous measurement [15]) gives $E_{max}$=1.699±0.005 eV, $\Gamma_G$=422±5 meV, and $\Gamma_L$=492±7 meV. As shown in Fig. 5(a) and 5(b), the spectral changes observed in the concentrated salt solutions can be simulated quantitatively by postulating that both $S_{max}$ and $E_{max}$ linearly increase with the ionic strength whereas the spectral shape parameters $\Gamma_G$ and $\Gamma_L$ do not change with the salinity (a better agreement can be obtained by allowing 5% narrowing of the spectral line at high salt concentration, but this correction does not make much difference in the absorbance at 700 nm). The band shift is 21.5±0.2 meV (sodium sulfate) and 24.9±0.3 meV (sodium perchlorate) per 1 M ionic strength. The relative increase in $S_{max}$ (relative to water) is 0.054 (sodium sulfate) and 0.014 (sodium perchlorate) per 1 M ionic strength. Using the plots shown in Figs. 5(a) and 5(b), it is possible to fit the dependence of the 700 nm absorbance shown in Fig. 5(c).

The constancy of the band shape and linear increase in $E_{max}$ were observed both for aqueous iodide and for hydrated electron in these concentrated salt solutions. For iodide, the extinction coefficient at the band maximum does not change with the salinity because the band shape and the oscillator strength are constant. [29,30] If the same pertains to hydrated electrons (the oscillator strength for the $1s \rightarrow 2p$ band of the electron [33] and the molar extinction coefficient of the electron at $E=E_{max}$ are constant), the relative extinction coefficient $\varepsilon_{700}/\varepsilon_{700}(I=0)$ at 700 nm (vs. water) can be readily calculated from our data (these simulations are shown in Fig. 7(a) in section 3.5). It follows from this calculation that this quantity is 0.95 for 1.8 M $Na_2SO_4$ and 0.8 for 9 M $NaClO_4$. However, the constancy of the oscillator strength is not apparent from our results because $S_{max}$ actually *increases* with the salt concentration. It should be emphasized that the optical density in Figs. 4 and 5 depends not only on the electron absorptivity but also on the electron yield. The latter depends on (i) the energy absorbed per volume of the sample (which rapidly increases with the density $\rho$ of the sample) and (ii) radiolytic yield of electrons (whose dependence on the salt concentration is unknown). Simple calculations suggest that in 9 M (66.4 wt%) $NaClO_4$, ca. 60% of the absorbed dose is stopped by the salt; i.e., this is not, strictly speaking, "water" radiolysis. Even when only one of these two factors is taken into account, the increase in the $S_{max}$ is negated: e.g., for sodium sulfate, $S_{max}/\rho$ is nearly constant with the salt concentration.

Since these pulse radiolytic data alone were insufficient to substantiate our (very plausible) assumption that the oscillator strength for electron band does not depend on the ionic strength of the solution, we needed to determine the QY for electron formation that did not rely on the (unknown) electron absorptivity. From that measurement, the molar absorptivity of the electron in saline solutions can be estimated. This program is implemented in section 3.5.

### 3.4. Relative quantum yields for electron detachment by 248 and 193 nm light.

As explained in the previous section, the absolute extinction coefficient $\varepsilon_{700}$ of the electron in saline solutions is not known accurately and, therefore, in this section we will limit ourselves to determination of relative quantum yield $\Phi_{rel}$ given by

$$\Phi_{rel} = \phi/\phi(I=0) \times \varepsilon_{700}/\varepsilon_{700}(I=0) \tag{1}$$

where $I$ is the ionic strength. The quantity $\Phi_{rel}$ can be determined directly from the plots of end-of-pulse transient absorbance of the electron at 700 nm ($\Delta OD_{700}$) vs. the absorbed laser power $I_{abs}$. Typical plots for 38.5 mM sulfite and 2 mM iodide in $Na_2SO_4$ solution are shown in Fig. 6(a) and 6(b), respectively All of the photosystems exhibited linear dependencies of $\Delta OD_{700}$ vs. $I_{abs}$ so that the product $\phi \varepsilon_{700}$ can be readily obtained from the slopes of the corresponding plots. These $\phi \varepsilon_{700}$ products were then plotted vs. the ionic strength of the solution; all such dependencies were linear. Extrapolation of these linear dependencies to zero ionic strength gave the estimate for $\phi(I = 0)$ $\varepsilon_{700}(I = 0)$ by which $\Phi_{rel}$ (eq. (1)) was calculated. The estimates for $\phi(I = 0)$ that were obtained in this work using the known molar absorptivity of hydrated electron in dilute aqueous solutions were within 3% of the values reported in Table 1, ref. [1].

One of the photosystems studied, OH⁻/NaClO₄ excited by 193 nm light, needed special treatment, because some of the excitation light was absorbed by sodium perchlorate (Figs. 6(c) and 5S(a)). As explained in section 3.1, very few electrons are generated by photoexcitation of this anion, and this extra 193 nm absorbance has the effect of reducing the electron yield from hydroxide. To compensate for this effect, the absorbance $OD_{193}$ of the solution at 193 nm was determined and plotted as a function of hydroxide concentration (Fig. 5S(a)). These dependencies were linear and exhibited a positive offset at $[OH^-] = 0$ (due to the light absorbance by perchlorate), from which the fraction of photons absorbed by hydroxide alone can be estimated (as the ratio of hydroxide absorbance $OD_{193}(OH^-)$ to the total absorbance). The absorbed laser power $I_{abs}$ was corrected by this ratio giving the laser power $I_{abs}(OH^-)$ absorbed by hydroxide anions alone: $I_{abs}(OH^-) = I_{abs}(OD_{193}(OH^-)/OD_{193})$. As shown in Fig. 6(c), when $\Delta OD_{700}$ is plotted against this latter quantity, the points obtained for different hydroxide concentrations are on the same line, which justifies the whole approach.

Fig. 5S(b) gives the resulting plots of $\Phi_{rel}$ as a function of the ionic strength (the data for electron photodetachment from perchlorate itself [1] are also included). (This figure has been placed in the Supplement because it is very similar to Fig. 7(b) in the next section). It is seen that the decrease in $\Phi_{rel}$ is similar for all four of the CTTS anions being studied (varying from 0.06 to 0.1 per 1 M of the ionic strength) in either sodium sulfate and sodium perchlorate solutions. Examination of Fig. 5S(b) suggests that $\Phi_{rel}$ decreases mainly due to a decrease in the *absolute* QY for electron photodetachment; the change in the molar absorptivity of the hydrated electron is relatively unimportant. Indeed, in 9 M NaClO$_4$, $\Phi_{rel}$ for iodide is ca. 20 times lower than in water; whereas the estimated decrease in $\varepsilon_{700}$ is just 20% (section 3.3). For 1.8 M Na$_2$SO$_4$, $\Phi_{rel}$ is 0.4 whereas the estimated decrease in $\varepsilon_{700}$ is just 5%. As demonstrated in section 3.5, the changes in the electron absorptivity at 700 nm are as small as suggested by our pulse radiolysis data, and the decrease in the absolute QY for electron photodetachment is as large as suggested by the data of Figs. 5 and 7(a).

It should be stressed that we are not first to look for the ionic strength effect on the electron yield. [8,34,35] These previous studies were excluded from our consideration so far because we believe that the data were compromised by inadequate choice of anion photosystems. In particular, Jortner et al. [34] claimed the absence of the effect for 254 nm photoexcitation of aqueous iodide. Specifically, they looked for an effect of addition of 2.5 M and 5 M HCl on the evolution of H$_2$ from a solution that contained 0.15 M iodide and 9 M LiCl. In a different experiment, the effect of addition 1 M KBr on the formation of iodine in N$_2$O-saturated solution containing 0.15 M KI and 1.8 M H$_2$SO$_4$ was sought. It is difficult to interpret these data since in both of these cases the reference systems were also concentrated solutions. Furthermore, under the experimental conditions, Cl$^-$ and Br$^-$ reacted with iodine atoms and hydrated electrons reacted with hydronium ions on a subnanosecond time scale, i.e., the geminate dynamics were perturbed in a major way. Ohno [35] observed that in 254 nm photoexcitation of hexacyanoferrate(II), the electron yield estimated from N$_2$ evolution in N$_2$O saturated solution decreased linearly with the concentration of sodium sulfate and sodium perchlorate (< 1 M) added to the reaction mixture (19% and 32% decrease per 1 M of ionic strength, respectively). This decrease is much greater than shown in Fig. 7(b), however, Fe(CN)$_6^{4-}$ is known to strongly associate with 1-3 alkali cations, [8] even in dilute solutions, [36] and in that propensity it is different from the anions studied herein. Furthermore, the absolute QYs obtained for Fe(CN)$_6^{4-}$ by Ohno [35] are 2 times lower than any other published estimate for this quantum yield.

### 3.5. Quantum yields for electron and $I_2^-$ formation in 248 nm photolysis of iodide.

The iodine atom generated from iodide in the course of electron photodetachment readily reacts with the iodide by hemicolligation (rxn. 3), forming $I_2^-$

$$I^- \xrightarrow{h\nu} I^\bullet + e_{aq}^- \qquad (2)$$

$$I^\bullet + I^- \longrightarrow I_2^- \qquad k=(1.2-1.3)\times 10^{10} \text{ M}^{-1}\text{ s}^{-1} \text{ [37]} \qquad (3)$$

This radical anion has almost no absorbance at 590 nm and absorbs strongly at 400 nm and 720-750 nm (the estimates for $\varepsilon_{400}$ range from 8300 to 15600 M$^{-1}$ cm$^{-1}$ [37] with 10000 M$^{-1}$ cm$^{-1}$ as the preferred value [37]; the visible absorption band is ca. 4 times weaker). Both of these absorption bands are intramolecular in origin; [38] neither the absorption spectrum nor the oscillator strength for this radical anion are expected to change with the water salinity, and determination of the concentration of $I_2^-$ from the transient absorbance is straightforward. In the absence of cross recombination, the yield of $I_2^-$ equals the yield of iodine atoms which in turn equals the electron yield. Combining this latter yield and the electron absorbance (determined under the identical photoexcitation conditions), one obtains the molar absorptivity of the electron. In practice, since the absorptivity of $I_2^-$ in water is estimated using the absorbance of electron in water as a reference, the quantity determined is the ratio $\varepsilon_{700}/\varepsilon_{700}(I=0)$ of the extinction coefficients for the electron (which is needed to separate the first factor in eq. (1)).

The experimental approach followed that given in section 4.2.3, ref. [1]. The 2 mM NaI solution was saturated either by $N_2$ or by $CO_2$. The latter solute served as an electron scavenger that "removed" the 700 nm absorbance of the electron in < 100 ns. The formation of $I_2^-$ can be observed either at 700 nm (where this relatively slow formation kinetics overlaps in time with the fast decay kinetics of the electron, see Fig. 6S and also Fig. 5 in ref. [1]) or at 400 nm (where the electron absorbance is negligible). We found that the ratio of the 700 nm and 400 nm absorbances at 0.2-1 μs (where the $I_2^-$ concentration reaches the maximum) was 0.3 regardless of the ionic strength of the solution; i.e., the shape of the absorption spectrum for $I_2^-$ indeed does not change with the salinity. At any given excitation light fluence, the maximum 700 nm absorbance from the electrons (in the $N_2$ saturated solution) and $I_2^-$ anions (in the $CO_2$ saturated solution) were in the same ratio, which suggests that there is very little decay of iodine atoms and $I_2^-$ anions via reaction with $CO_2^-$ anions (that migrate and react much slower than hydrated electrons). That much can also be deduced by comparing the decay kinetics of 700 nm absorbance in these $N_2$ and $CO_2$ solutions: the decay of $I_2^-$ in the $I^\bullet/I_2^- + CO_2^-$ mixture is almost an order of magnitude slower than the decay of electrons in the $I^\bullet/I_2^- + e_{aq}^-$ mixture (Fig. 6S). To reconcile the QYs determined for $I_2^-$ and the electron in dilute aqueous solutions, an extinction coefficient of 2290 $M^{-1}$ $cm^{-1}$ was assumed for the former species at 700 nm (in reasonable agreement with 2500 $M^{-1}$ $cm^{-1}$ at 720 nm given by Elliot and Sopchyshyn [37]). Using this estimate and products $\phi\varepsilon_{700}$ for $I_2^-$ and $e_{aq}^-$ formation that were determined as explained in sections 2 and 4.4, the ratio $\varepsilon_{700}/\varepsilon_{700}(I=0)$ was obtained (Fig. 7(a)). In full agreement with section 3.3, the molar absorptivity $\varepsilon_{700}$ of the electron in $Na_2SO_4$ is nearly constant with the ionic strength of the solutions; for $NaClO_4$, this molar absorptivity decreases by 20% in the 9 M $NaClO_4$ solution. Though for the latter salt the $\varepsilon_{700}/\varepsilon_{700}(I=0)$ dependence shown in Fig. 7(a) is more curved than calculated, the accuracy of the latter measurement is perhaps lower, and we used the calculated data to obtain the ratios $\phi/\phi(I=0)$ given in Fig. 7(b). Since the decrease in the electron absorptivity is relatively small, the two plots (Figs. 5S(b) and 7(b)) look quite similar.

To summarize these QY data, the quantum yield of electron photodetachment rapidly decreases with increasing ionic strength of the solution. The magnitude of the decrease in the QY per 1 M ionic strength is similar for all of the photosystems studied. Note that the decrease in the molar absorptivity of CTTS anions with the ionic strength has already been factored into the calculation of the QYs since we directly measured the absorption of the excitation light by these solutions. The concomitant decrease in the molar absorptivity of the electrons has also been taken into account.

The QYs determined using flash photolysis are quantum efficiencies for the formation of *free* electron that escaped recombination with its geminate partner. This quantity is given by the product of a prompt QY for the electron formation (which depends on the details of dissociation of a short-lived CTTS state on subpicosecond time scale) and a fraction $\Omega_\infty$ of electrons that escape geminate recombination on a sub-nanosecond time scale. It is not obvious from the data of Fig. 7(b), which one of these two factors is responsible for the observed decrease in the (free) electron yield with the increased ionic strength. The previous observation of an ionic strength effect on electron detachment from $Fe(CN)_6^{4-}$, by Ohno, [35] was interpreted by the author as the effect of solvent cage constriction around the CTTS state; in other words, the decrease in the free electron yield was thought to result from the decrease in the prompt electron yield attained in the course of this CTTS state dissociation. In the next section, ultrafast spectroscopy is used to demonstrate that the ionic strength effect is dynamic in origin and develops on the time scale of hundreds of picoseconds; the prompt electron yield does not change with the salt concentration.

### *3.6. Ultrafast 200 nm pump - 800 nm probe spectroscopy of anions in concentrated solutions.*

### *3.6.1. Bromide.*

Bromide has an absorption maximum at 200 nm, [29] and blue shift of the absorption band with the ionic strength has relatively little effect on the molar absorptivity of this anion at the photoexcitation wavelength (as compared, for example, with iodide at 248 nm). As explained in section 3.1, due to the large absorptivity of this anion at 200 nm (10000 $M^{-1}$ $cm^{-1}$), photoexcitation of sulfate and perchlorate is negligible in the 20-80 mM NaBr solutions that were used. No change in the electron dynamics were observed with bromide concentration (< 100 mM) in these solutions (Fig. 7S). Another advantage of using bromide is that the decay kinetics of electron generated by 200 nm and 225 nm photoexcitation of this anion were found to be identical, i.e., it appears that direct ionization with the formation of a conduction band electron does not occur for bromide across the low-energy CTTS subband (see Part III of this series [4]).

Typical 800 nm transient absorption kinetics obtained from 30 mM bromide in sodium perchlorate solutions are given in Fig. 8(a). The data for sulfate solutions are shown in Figs. 7S and 8S. The initial very rapid kinetics ($t < 3$ ps) are from electron thermalization and solvation. For higher concentrations of NaClO$_4$, these kinetics become progressively faster and the optical density at 5-10 ps systematically decreases. Both of these trends are due to shifting of the absorption band of thermalized electron to the blue (section 3.3). The decrease in the prompt signal from presolvated electron with NaClO$_4$ concentration is relatively small (because the change in bromide absorptivity is relatively small and the absorptivity of this presolvated species changes little with the ionic strength). In water, the maximum of the electron band is ca. 719 nm, [15,33] which is close to our probe wavelength (800 nm). In 9 M NaClO$_4$, this band maximum shifts to 636 nm. From previous studies of electron thermalization in water photoionization [39,40,41] and iodide photoexcitation [3a] it is known that the band maximum of presolvated electron systematically shifts to the blue in the first 250 fs. After the first 300-500 fs, this maximum reaches its equilibrium position [3a,39,41] and further spectral evolution is due to narrowing of the band to the red of the band maximum and widening of the band to the blue of the band maximum [40]. The greater is the observation wavelength with respect to this maximum, the greater is the reduction in the electron absorbance during the thermalization process and the faster is the thermalization kinetics. This can be demonstrated by changing the observation wavelength [40] or by increasing the temperature of the solution and probing at a fixed wavelength (the spectrum of hydrated electron shifts to the red with increasing temperature, and the effect is exactly opposite to that shown in Fig. 8(a)) [41,42] The decrease of the 800 nm absorbance at the end of the thermalization process (at 5-10 ps) with [NaClO$_4$] reflects the decrease in the electron absorptivity; when the latter is taken into account using the data of section 3.3, there is almost no variation in the prompt yield of the thermalized electron.

In Fig. 8(b), the kinetics shown in Fig. 8(a) were normalized at 5 ps. It is seen that these normalized kinetics begin to diverge only after the first 40-50 ps. At 500 ps, this divergence is very large: the probability $\Omega(t)$ of the electron survival rapidly decreases at higher ionic strength. The same behavior was observed for sodium perchlorate and sodium sulfate solutions of bromide. To a first approximation, these kinetics are biexponential [2,3,4,5], and the escape probability $\Omega_\infty$ for $t \to \infty$ can be readily estimated from the corresponding least squares fits (as explained in section 3.6.3, a more complex type of fit has actually been used, but the main result does not depend on the specific prescribed profile of the kinetics). These escape probabilities are plotted vs. the ionic strength of the sodium sulfate and the sodium perchlorate solutions in Fig. 9(a). The survival probability of the geminate pair decreases linearly with the ionic strength with a slope of ca. 9 % per 1 M. This slope is close to the slope of 8-12 % per 1 M observed for the decrease in QY of free electron in 248 nm photoexcitation of iodide and sulfite (Fig. 7(b)). Thus, the answer to the question formulated at the end of section 3.5 is that *the QY of free electron decreases as a function of the ionic strength due to the decrease in the fraction of electrons that escape geminate recombination; the prompt yield of thermalized electrons does not change with the addition of salts*. Interestingly, the transformations of the kinetics with the increasing ionic strength shown in Fig. 8(b) are exactly opposite to the transformations observed for the decay kinetics in 200 nm photoexcitation of hydroxide with increasing temperature. An increase in the temperature red-shifts the spectra of halide/pseudohalide anions and the hydrated electron and increases $\Omega_\infty$; the main difference between the effect of the salt addition and the effect of the temperature increase is that for the latter the normalized kinetics diverge on the time scale of tens rather than hundreds of picoseconds. Still, the overall behavior is strikingly similar: *the effect of increasing the ionic strength largely resembles the effect of lowering the solution temperature*.

We will return to the interpretation of the kinetics in Figs. 8(b) and 8S in section 3.6.3. For now, it suffices to note that the ionic strength effect on the geminate recombination is slow to develop: for electrons photodetached from bromide it takes 300-500 ps for this effect to develop fully. As shown in the next section, for other photosystems this time lag is even longer.

### 3.6.2. Sulfite and hydroxide.

Fig. 10(a) shows the decay kinetics of electrons obtained in 200 nm photoexcitation of 40 mM sulfite in concentrated $Na_2SO_4$ solution (non-normalized kinetics are shown in Fig. 9S). As in the case of bromide (section 3.6.1), the photoinduced optical density at the end of a 200 nm pulse changes little as a function of salinity, and the thermalization kinetics follow the same transformations as those observed for bromide in NaClO$_4$ solutions (albeit on a smaller scale, due to a smaller blue shift of the hydrated electron band in the sodium sulfate solution, Fig. 4). When these kinetics are normalized at 5-10 ps (at the end of the thermalization kinetics) the "tails" of these kinetics ($t$ >100 ps) barely diverge, though one can see that the escape of electrons in water is more efficient than this escape in 1.83 M Na$_2$SO$_4$ solution. Much the same behavior is observed for hydroxide (Fig. 9S(b)): the changes in the kinetics are confined to thermalization dynamics; the normalized kinetics after 10 ps are nearly the same, with barely perceptible divergence after the first 200 ps. The lack of an ionic strength effect on these $t$<500 ps kinetics is not completely unexpected: Crowell and coworkers [5,6] have studied the effect of addition 1-10 M KOH on the electron and OH radical dynamics in 200 nm photoexcitation of hydroxide. No effect was observed. Thus, for hydroxide and sulfite, not only there is no effect of the ionic strength on the prompt yield; there appears to be almost no decrease in the survival probability of the geminate pair in the first 500 ps after the electron detachment.

At first glance, this observation appears to contradict the QY estimates of section 3.5 in which a large effect of the ionic strength on the *free electron* yield was observed for 193 nm photoexcitation of hydroxide and 248 nm photoexcitation of sulfite. Given that all of these photosystems exhibit very similar behavior, we find it unlikely that the electron dynamics in 200 nm photoexcitation of hydroxide follow an entirely different set of rules than the electron dynamics in 193 nm photoexcitation of the same anion. Apparently, what distinguishes hydroxide and sulfite from bromide is that the time scale on which the geminate dynamics diverge (eventually approaching similar $\Omega_\infty$) is even *longer:* the full effect is attained on the time scale of a few nanoseconds.

*3.6.3. Kinetic simulations for bromide.*

As stated in the Introduction, electron dynamics observed in the previous studies of electron photodetachment from monovalent aqueous anions broadly conformed to the MFP model. One particular realization of this model is the semianalytical theory of Shushin [7]. We refer the reader to refs. [5] and [7] for rigorous formulation of that model. Very briefly, Shushin [7] showed that under most realistic conditions, the time dependent survival probability $\Omega(t)$ of a geminate pair is not sensitive to the radial profile $U(r)$ of the MFP *per se*. Rather, this probability depends on a few global parameters that can be analytically derived from this potential. These parameters are the Onsager radius $a$ (at which the potential equals the Boltzmann energy, $U(a) = -kT$), the first order rates $W_r$ and $W_d$ for the recombination and dissociation from the potential well ($r<a$), respectively (so that $W = W_d + W_r$ is the total rate of the decay of pairs inside the potential well). From these three parameters and the mutual diffusion coefficient $D$ of the pair, two dimensionless parameters can be obtained: the survival probability $p_d = W_d / W \ (= \Omega(t \to \infty))$ of the geminate pair generated inside the potential well and $\alpha = a p_d \sqrt{W/D}$, which characterizes the rate with which the short term exponential kinetics $\Omega(t) \propto \exp(-Wt)$ transforms into long term kinetics tailing as $p_d(1 + a/\sqrt{\pi D t})$. The overall kinetics is given by a rather complex formula; the important point is that all experimental traces observed so far can be fit by this formula, or its minor modification [4,5] that provides for the possibility that some geminate pairs are initially generated beyond the Onsager radius of the MFP.

The traces for Br$^-$ shown in Fig. 8(b) and 8S can also be simulated using Shushin's theory equations. Since for bromide no difference was observed between the kinetics obtained using 200 nm and 225 nm photoexcitation, we assumed that addition of salts does not change the initial distribution of electrons around bromine atom and assumed that this initial distribution is entirely confined within the potential well.

Fig. 8S shows a global fit of normalized kinetics for the Br$^-$/Na$_2$SO$_4$ photosystem using Shushin's theory equations. In these least squares fits, parameters $W^{-1}$ and $\alpha$ were assumed to be constant (19.5 ps and 0.552, respectively); only the survival probability $\Omega_\infty = p_d$ was allowed to change (Fig. 9(a) gives the optimum values for this parameter as a function of the ionic strength). While this way of fitting the kinetic data obviously works, all it really conveys is that the change in $p_d$ is the main cause for the kinetic transformations shown in Figs. 8(b) and 8S. Indeed, since $\alpha$ and $p_d$ explicitly depend on *W*, it is difficult to explain how the constancy of *W* and $\alpha$ can be reconciled with a large change in $p_d$, even if the fit quality is good.

For sodium perchlorate solutions, the range of the kinetic change is greater and it becomes apparent that this constancy with the ionic strength is not maintained: it is impossible to find optimum parameters *W* and $\alpha$ for which a change in $p_d$ alone accounts for the observed kinetic transformations. Assuming constancy of the Onsager time $\tau_c = a^2/D$ with the ionic strength, rate constants $W_d$ and $W_r$ can be determined from the least squares fits for individual kinetic traces in Fig. 8(b). As shown in Fig. 9(b), these optimum rate constants, the decay constant *W*, and the survival probability $p_d$ all decrease linearly with the ionic strength. Since $W_d$ and $W_r$ are complex functions of the potential *U(r)*, it is impossible (in the framework of Shushin's theory) to pinpoint what specific modifications of the MFP are required to obtain the changes shown in Fig. 9(b). Provided that the Onsager radius of the MFP does not change with ionic strength, only lowering of the reduced potential energy *U(r)/kT* (i.e., making the MFP more attractive) can decrease both of these two rate constants. Thus, the most general conclusion that can be made from fitting these kinetics using the MFP model in its formulation by Shushin is that addition of salts strengthens the attraction between the electron and bromine atom. This partially accounts for the qualitative similarity of the temperature effect and the ionic strength effect: increase in the temperature *decreases* this reduced potential energy *U(r)/kT* and increases $W_d$ and $W_r$.

## 4. Discussion.

The following picture emerges from the results of section 3: Addition of molar concentration of salts to aqueous solutions of CTTS anions does not change the prompt quantum yield for the formation of hydrated electron; rather, it decreases the survival probability of the geminate electron. Monovalent and polyvalent anions exhibit exactly the same behavior; the anticipated difference in the effect of ion screening on the two types of geminate pairs originating from these anions (section 1) was not observed. For the same ionic strength, the decrease in the survival probability $\Omega_\infty$ of hydrated electron is similar for all anions (Figs. 7(b) and 9(a)), but the rate by which the full ionic strength effect is attained varies considerably among the photosystems. For one photosystem only, bromide anion photoexcited by 200 nm light, were we able to observe the limiting value of the survival probability in 500 ps (which was our observation window, Fig. 8(b)). For bromide, the kinetics obtained at different ionic strengths started to diverge after 50 ps, when the fast exponential component was over, and the effect developed fully on the time scale of hundreds of picoseconds, when the decay of the electron was controlled by diffusional migration of the geminate partners to the bulk and their re-encounters. It is safe to conclude that the ionic strength effect develops slowly and requires many such re-encounters. Analysis of the kinetics for $Br^-/NaClO_4$ photosystem in the framework of the MFP well model of Shushin [7] given in section 3.6.3 suggests that the changes in the kinetics can be rationalized in terms of the increased attraction between the geminate partners in high ionic strength solutions.

Some of these observations broadly agree with the anticipated effect of charge screening. For polyvalent anions, a decrease in the long-range repulsion between the geminate partners facilitates their recombination thereby decreasing the survival probability of the electron. Because this repulsion is already screened by water, the effect can be observed only at high concentration of the screening ions and only when the partners are close; thus, many reencounters are needed. Since photoexcited polyvalent anions (unlike monovalent anions) yield a broad distribution of separation distances between the electron and its geminate partner (presumably, due to the occurrence of direct ionization [4,11,12b]), the average distance between these partners is always very long (2-3 nm) and the screening (which is a relatively short-range effect) takes a longer time to perturb the geminate dynamics.

Other observations are more difficult to rationalize. For monovalent anions, the MFP potential is attractive and thought to originate through weak electrostatic interactions between the geminate partners. Screening of these interactions would make the attractive force weaker thereby reducing the efficiency of recombination and increasing the survival probability. This, apparently, does not happen. The most difficult observation to accommodate is that photosystems having different overall MFPs all show similar relative decrease in the survival probability when this potential is modified by intervening ions. In principle, the decrease in the survival probability for pairs originating from monovalent anions can be "explained" within the MFP model with ion screening. E.g., it can be postulated that for geminate pairs derived from monovalent anions, short-range weak attraction gives way to long-range weak repulsion at greater distances between the geminate partners (as is likely to be the case for polyvalent anions), and it is this repulsion that is screened by the ions. Alternatively, one can postulate that the reactivity of close pairs systematically increases with the ion concentration (e.g., due to pairing of the electron with a cation) and this effect counteracts the weakening of the attractive potential between the partners (though this rationale is incompatible with the $W_{r,d}$ data of Fig. 9(b)). In our viewpoint, such modifications are both forced and misguided because the similarity of the ionic strength effect for various photosystems emerges only as a consequence of fine tuning of the model parameters in each particular case. Can something other than the ion screening cause the ionic strength effect? Below, we consider three alternate scenarios in detail.

The first scenario is that the ionic strength effect originates through narrowing of the initial distribution of the electrons. The idea is that an addition of salt causes a blue shift in the absorption spectrum of an anion and therefore is equivalent to photoexcitation of this anion (in water) at a lower energy. Since it is known that for some anions an increase in the excitation energy causes more electrons to escape (due to the broadening of the electron distribution via enhanced direct ionization), [4,11] the salt-induced blue shift causes the opposite effect. Though this scenario cannot be excluded for sulfite [4] and hydroxide [4,5], it can be completely excluded for iodide and bromide because for these two halide anions no wavelength dependence was observed to the red of their CTTS band maxima. [3,4,11]

The second scenario is that the effect originates via a purely kinematic effect, such as an increase in the solution viscosity and/or the ionic atmosphere drag on the electron diffusion. The drag of a slowly-migrating ionic atmosphere on the rapidly-migrating ion is a well known effect: the ionic atmosphere needs to readjust itself as the fast ion (e.g., hydrated electron) migrates. In the Debye-Hückel theory of aqueous electrolytes (applicable to $I < 0.1$ M), the relaxation time for the ionic atmosphere is $55/I$ ps, i.e., at the upper limit of validity of this theory, it takes the ionic atmosphere ca. 500 ps to relax [5]. Perhaps, this relaxation time continues to decrease in concentrated solutions (although this trend is opposed by slowing of the ion diffusion (Fig. 1S(b) and ion pairing (Fig. 2S(b)). Thus, it is possible that the diffusion coefficient of the electron systematically decreases with delay time $t$ in the observation window of our ultrafast kinetic experiments. However, one would expect that the drag on the electron diffusion develops on the same time scale in all photosystems. Almost complete lack of kinetic variation with the ionic strength for some photosystems (e.g., for hydroxide and sulfite, see refs. [5], [6] and section 3.6.2) suggests that this effect is relatively minor, at least on a sub-nanosecond time scale. Furthermore, while a time-dependent (or decreased) electron diffusivity would change the temporal profile of the kinetics, it cannot change the survival probability: in the MFP model, the latter depends on the potential only. Thus, this scenario should also be rejected.

The third and last scenario is ion association. Before proceeding any further, let us consider the existing explanations for salt induced blue shift in the absorption spectra of halide (e.g., iodide) anions and the hydrated electron. Both of these species exhibit very similar blue shifts for a given salt concentration (cf. Figs. 2(b) and 5(b)). Since the original suggestion by Stein and Treinin [30] most authors accounted for this effect by constriction of the solvation cavity around the anion/electron (some more exotic explanations have been considered by Blandamer and Fox [1]), though the cause for this constriction has never been specified. The blue shift of the solvated/trapped electron spectrum is observed both in liquid and solid vitreous solutions, such as alkaline glasses [43]. For the latter, electron spin echo envelope modulation (ESEEM) spectroscopy provides direct evidence for the inclusion of sodium cation in the first solvation shell of the trapped electron through the observation of dipole and Fermi coupling of the unpaired electron to the spin of $^{23}$Na nucleus [44]. In alkaline earth salt aqueous glasses [43a] and THF solutions, [43b] electrons associated with one or several cations can be distinguished by optical spectroscopy [43a]. In most of these glasses, the visible band coexists with an IR band that gradually shifts to the blue; for alkaline glasses the visible band itself shifts to the blue on a sub-ms timescale. [43a] Many authors speculated that the cations are actively involved in the trapping site of the electron, and specific "ionic lattice trap" models detailing this involvement were formulated by Hilczer et al. [45] and Bartzak et al. [46]. In short, there is much evidence pointing to structural involvement of alkali cations in electron dynamics observed for low temperature glassy solutions.

We suggest that the same process occurs in liquid aqueous solutions at room temperature: sodium cations are included in the solvation shell of the electron. The thermalization kinetics given in section 3.6.1 indicate that in concentrated salt solutions, the electron spectra blue shift on a time scale similar to that of electron thermalization in pure water. This means that the postulated association of sodium cation(s) with the electron occurs very rapidly, in less than 5 ps. Assuming a diffusion controlled association of the sodium cation and the electron and taking the diffusion coefficients for the electron (in water) and the sodium cation as $4.9 \times 10^{-5}$ and $1.2 \times 10^{-5}$ cm$^2$/s, respectively, the association in 1 M NaClO$_4$ would take ca. 45 ps, whereas the blue shift in the electron spectrum develops in < 5 ps. This estimate suggests that the cation associated with the electron originates either from the dense part of the ion atmosphere around the parent anion or from a cation that is associated with this anion (thereby causing the blue shift of the anion spectrum). That the latter association occurs to some degree (at least, for polyvalent anions) via the formation of solvent-separated ion pairs, follows from many independent observations (e.g., see section 3.1) which point to such pairs as the main entity in concentrated solutions. The association of $e^-_{aq}$ and Na$^+$ has been previously suggested by Gauduel and coworkers to account for electron thermalization and subpicosecond dynamics of geminate pairs generated in 2-photon excitation of aqueous chloride. [47]

Thus, we suggest that in concentrated salt solutions, both the anion and the electron are associated with alkali cations and that the cation associated with the parent anion (or included in the inner part of the ionic atmosphere around this parent anion) is passed to the electron in the course of CTTS state dissociation on subpicosecond time scale, possibly, in a manner envisioned by Gauduel and coworkers. [47] The fraction of associated anions and electrons rapidly increases with the ion concentration. This association makes the short-range potential more attractive, for all anions, including polyvalent anions. A possible cause for this effect is the increased attraction between the radical/atom and hydrated electron separated by a cation; rather than screening the weak attractive forces, such a cation might bind these species together by excluding the screening water dipoles.

Recently, Lenchenkov et al. studied the effect of addition of 0.04-4 M KBr on the electron dynamics in 255 nm photoexcitation of ferrocyanide [8]. A 4% decrease per 1 M of KBr in the survival probability of geminate pairs (extrapolated from the kinetic data obtained for $t$<400 ns) was observed. Both ionic atmosphere screening and ion association were considered, and the latter was concluded to be the main cause of the observed effect. Specifically, association of ferrocyanide with 1-3 potassium cations was suggested to reduce the Coulomb repulsion between the geminate partners. [8] As seen from the results of the present work, the ionic strength effect occurs even for geminate pairs in which no such repulsion exists. Still we believe that the root cause of this effect is the same for all photoexcited anions: the short-range MFP (which may or may not include Coulomb repulsion) is modified by the ion association.

While this picture is incomplete, it contains a verifiable rationale: the ionic strength effect originates from the modulation of short-range attraction between the geminate partners rather than long-range ion screening. A modification of short-range MFP for the $\left( X^{(n-1)-\bullet}, e_{aq}^{-} \right)$ pair due to the involvement of alkali cation(s) in the solvation of the parent anion ($X^{n-}$) and the electron $e_{aq}^-$ is postulated. For reasons not entirely understood, this association makes the attractive MFP interaction stronger thereby causing all of the changes observed in section 3.6. Only detailed molecular dynamics simulation of this charge separation process (along the lines of refs. [10,13,48,49,50]) can shed more light on the mechanism. The first step in this direction has already been taken by Boutin and co-workers who used the umbrella sampling approach of Borgis and Staib [49] to obtain the MFP for interaction of hydrated electron and $Na^+$. This simulations suggests the existence of a contact $(Na^+, e_{aq}^-)$ pair with the average distance of ca. 0.2 nm and the binding energy of 3-4 $kT$.

**5. Conclusion.**

It is shown that in concentrated solutions of $NaClO_4$ and $Na_2SO_4$, the quantum yield for (free) electron photodetachment from CTTS anions (such as $I^-$, $OH^-$, $ClO_4^-$, and $SO_3^{2-}$) decreases linearly by 6-12% per 1 M ionic strength. In 9 M sodium perchlorate solution, this QY decreases by roughly an order of magnitude. Ultrafast kinetic studies on 200 nm photon induced electron detachment from $Br^-$, $HO^-$ and $SO_3^{2-}$ suggest that the prompt yield of thermalized electron in these solutions does not change; the ionic strength effect wholly originates in the decrease in the survival probability of geminate pairs. Kinetically this effect manifests itself in much the same way as the temperature effect studied previously, [5] that is, addition of salt and decrease in temperature cause similar changes in the electron dynamics. In this respect, the ionic strength effect on the QY follows the same pattern as the shift in the absorption spectra of the CTTS anions and the hydrated electron itself. [1,31,33]

Within the framework of the recently proposed MFP model of charge separation dynamics in CTTS photosystems, [2,5] the observed changes can be explained by an increase in the short-range attractive potential between the geminate partners (possibly, due to the association of sodium cation(s) with the electron and the parent anion); ion screening plays minor role. The precise mechanism for modification of the MFP by these cations is presently unclear, although electron thermalization kinetics for the bromide photosystem suggest that the sodium cation associated with the parent $Br^-$ anion is passed to the photodetached electron. [47] Further advance in the understanding of the ionic strength effect is predicated on the development of a molecular dynamics model that directly includes the interactions of the geminate partners with cations in the ionic atmosphere.

**6. Acknowledgement.**

We thank Profs. S. E. Bradforth of USC, B. J. Schwartz of UCLA and Dr. S. V. Lymar of BNL for many useful discussions. The research at the ANL was supported by the Office of Science, Division of Chemical Sciences, US-DOE under contract number W-31-109-ENG-38.

***Supporting Information Available:*** A PDF file containing Figs. 1S to 9S with captions. This material is available free of charge via the Internet at http://pubs.acs.org.

## References.

1. Sauer, Jr., M. C.; Crowell, R. A.; Shkrob, I. A., submitted to *J. Phys. Chem. A*; Part I of this series (jp049722t); preprint available on http://www.arXiv.org/abs/physics/0401080

2. Kloepfer, J. A.; Vilchiz, V. H.; Lenchenkov, V. A.; Chen, X.; Bradforth, S. E. *J. Chem. Phys.* **2002**, *117*, 776.

3. (a) Vilchiz, V. H.; Kloepfer, J. A.; Germaine, A. C.; Lenchenkov, V. A.; Bradforth, S. E. *J. Phys. Chem. A* **2001**, *105*, 1711; (b) Kloepfer, J. A.; Vilchiz, V. H.; Germaine, A. C.; Lenchenkov, V. A.; Bradforth, S. E. *J. Chem. Phys.* **2000**, *113*, 6288; (c) Kloepfer, J. A.; Vilchiz, V. H.; Lenchenkov, V. A.; Bradforth, S. E., in *Liquid Dynamics: Experiment, Simulation, and Theory* (2002), Vol. 820, pp. 108; Kloepfer, J. A.; Vilchiz, V. H.; Lenchenkov, V. A.; Bradforth, S. E. *Chem. Phys. Lett.* **1998**, *298*, 120.

4. Lian, R.; Oulianov, D. A.; Crowell, R. A.; Shkrob, I. A.; Chen, X.; Bradforth, S. E. *in preparation*, to be submitted to *J. Phys. Chem*. (Part III of the series)

5. Lian, R.; Crowell, R. A.; Shkrob, I. A.; Bartels, D. M.; Chen, X.; Bradforth, S. E. *J. Chem. Phys.;* jcpid A4.01.130; preprint available on http://www.arXiv.org/abs/physics/0401071.

6. Lian, R.; Crowell, R. A.; Shkrob, I. A.; Bartels, D. M.; Oulianov, D. A.; Gosztola, D. submitted to *Chem. Phys. Lett.*; preprint available on http://www.arXiv.org/abs/physics/0401057.

7. Shushin, A. I. *J. Chem. Phys*, **1992**, *97*, 1954; *Chem. Phys. Lett.* **1985**, *118*, 197, *J. Chem. Phys.* **1991**, *95*, 3657.

8. Lenchenkov, V. A.; Chen, X.; Vilchiz, V. H.; Bradforth, S. E. *Chem. Phys. Lett.* **2001**, *342*, 277.

9. Bradforth, S. E., *private communication.*

10. Borgis, D.; Staib, A. *J. Chem. Phys.* **1996**, *104*, 4776 and 9027.

11. Chen, X.; Kloepfer, J. A.; Bradforth, S. E.; Lian, R.; Crowell, R. A.; Shkrob, I. A. *(in preparation).*

12. (a) Martini, I. B.; Barthel, E. R.; Schwartz, B. J. *J. Am. Chem. Soc.* **2002**, 124, 7622; Barthel, E. R.; Martini, I. B.; Schwartz, B. J. *J. Chem. Phys.* **2000**, *112*, 9433; Barthel, E. R.; Martini, I. B.; Keszei, E.; Schwartz, B. J. *J. Chem. Phys.* **2003**, *118*, 5916; (b) Barthel E. R.; Schwartz, B. J. *Chem. Phys. Lett.* **2003**, *375*, 435.

13. Smallwood, C. J.; Bosma, W. B.; Larsen, R. E.; Schwartz, B. J. *J. Chem. Phys.* **2003**, *119*, 11263.

14. Elliot, A. J.; Ouellette, D. C.; Stuart, C. R. *The Temperature Dependence of the Rate Constants and Yields for the Simulation of the Radiolysis of Heavy Water*, AECL report 11658 (AECL Research, Chalk River Laboratories, Chalk River, Ontario, Canada, 1996).

15. Jou, F.-Y.; Freeman, G. R. *Can. J. Chem.* **1979**, *57*, 591.

16. Shkrob, I. A.; Sauer, Jr., M. C. *J. Phys. Chem. A* **2002**, *106*, 9120.

17. Zaytsev, I. D.; Aseyev, G. G. *Properties of Aqueous Electrolytes* (CRC Press, Boca Raton, 2000), pp. 7 and 164.

18. *Solubilities of Inorganic and Organic Compounds. v. 1. Binary Systems. Part I.* Eds. Stephen, H.; Stephen, T. (Pergamon, New York, 1963).

19. Kohner, H. *Z. Phys. Chem. B* **1928**, *1*, 427.

20. *Handbook of Chemistry and Physics, 69th edition,* ed. Weast, R. C.; Astle, M. J.; Beyer, W. H. (CRC Press, Boca Raton, 1988).

21. ref. [17], pp. 924.

22. Daly, F. P.; Brown, C. W.; Kester, D. R. *J. Phys. Chem.* **1972**, *76*, 3664.

23. Buchner, R.; Capewell, S. G.; Hefter, G.; May, P. M. *J. Phys. Chem. B* **1999**, *103*, 1185.

24. Fischer, F. H. *J. Solution. Chem.* **1975**, *4*, 237; Gilligan III, T. J.; Atkinson, G. *J. Phys. Chem.* **1980**, *84*, 208.

25. Weingärtner, H.; Price, W. E.; Edge, A. V. J.; Mills, R. *J. Phys. Chem.* **1993**, *97*, 6290.

26. Dainton, F. S.; Fowles, P. *Proc. Roy. Soc. A* **1965**, *287*, 312 and 295.

27. Iwata, A.; Nakashima, N.; Kusaba, M.; Izawa, Y.; Yamanaka, C. *Chem. Phys. Lett.* **1993**, *207*, 137.

28. Rabinovitch, E. *Rev. Mod. Phys.* **1942**, *14*, 112.

29. Blandamer, M. J.; Fox, M. F. *Chem. Rev.* **1970**, *70*, 59.

30. Stein, G.; Treinin, A. *Trans. Faraday Soc.* **1960**, *56*, 1393.

31. Fox, M. F.; Hayon, E. *J. Chem. Soc. Faraday Trans. I* **1976**, *72*, 1990.

32. Anbar, M.; Hart, E. J. *J. Phys. Chem.* **1965**, *69*, 1244.

33. Hart, E. J.; Anbar, M. *The Hydrated Electron* (Wiley-Interscience, New York, 1970).

34. Jortner, J.; Ottolenghi, M.; Stein, G. *J. Phys. Chem.* **1964**, *68*, 247.

35. Ohno, S. *Bull. Chem. Soc. Jpn.* **1967**, *40*, 1776 and 1770.

36. Cohen; S. R.; Plane, R. A. *J. Phys. Chem.* **1957**, *61*, 1096; Davies, C. W. *J. Am. Chem. Soc.* **1937**, *59*, 1760.

37. Elliot, A. J.; Sopchyshyn, F. C. *Int. J. Chem. Kinet.* **1984**, *16*, *1247*; Devonshire, R.; Weiss, J. J. *J. Phys. Chem.* **1968**, *72,* 3815; Zehavi, D.; Rabani, J. *J. Phys. Chem.* **1972**, *76*, 312.

38. e.g., Tasker, P. W.; Balint-Kurti, G. G.; Dixon, R. N. *Mol. Phys.* **1976**, *32*, 1651.

39. Lian, R.; Crowell, R. A.; Shkrob, I. A. (in preparation); abstract available in Abstracts of Papers, 226th ACS National Meeting, New York, NY, United States, September 7-11, 2003.

40. *see, for example* Hertwig, A.; Hippler, H.; Unterreiner, A.-N. J. Phys. Condens. Matter 2000, 12, A165; Pépin, C.; Goulet, T.; Houde, D.; Jay-gerin, J.-P. J. Phys. Chem. A **1997**, *101*, 4351; Laenen, R.; Roth, T.; Laubereau, A. *Phys. Rev. Lett.* **2000**, *85*, 50; Assel, M.; laene, R.; Laubereau, A. J. Phys. Chem. A **1998**, *102*, 2256; Shi, X.; Long, F. H.; Lu, H.; Eisenthal, K. B. J. Phys. Chem. **1996**, *100*, 11903.

41. Hertwig, A; Hippler, H.; Unterreiner, A.-N. *Phys. Chem. Chem. Phys.* **2002**, *4*, 4412.

42. Lian, R.; Crowell, R. A.; Shkrob, I. A.; Pommeret, S. (in preparation).

43. (a) Kroh, J. in *Pulse Radiolysis*, ed. Tabata, Y. (CRC Press, Boca Raton, 1990), Ch. 16, pp. 358; (b) Renou, F.; Archirel, P.; Pernot, P.; Levy, B.; Mostafavi, M. *J. Phys. Chem. A* **2004**, *108*, 987.

44. Dikanov, S. A.; Tsvetkov, Y. D. *Electron Spin Echo Envelope Modulation (ESEEM) Spectroscopy* (CRC Press; Boca Raton; 1992).

45. Hilcer, M.; Bartczak, W. M. ; Sopek, M. *Radiat. Phys. Chem.* **1985**, *26*, 693.

46. Bartzak, W. M.; Hilczer, M.; Kroh, J. Radiat. Phys. Chem. **1981**, *17*, 431 and 481.

47. For the ultrafast dynamics of the tentative $\left(Cl^{\bullet}, e_{aq}^{-}...Na^{+}\right)$ complexes and the discussion of possible mechanisms for rapid in-cage charge separation in concentrated $H^+$, $Li^+$, and $Na^+$ chloride solutions, see Gauduel, Y.; Gelabert, H.

*Chem. Phys.* **2000**, *256*, 333; Gelabert, H.; Gauduel, Y. *J. Phys. Chem.* **1996**, *100*, 13993; Gauduel, Y.; Gelabert, H.; Ashokkumar, M. *Chem. Phys.* **1995**, *197*, 167; similar mechanisms were also suggested by these workers for divalent cations, such as $Mg^{2+}$ and $Cd^{2+}$, see Gauduel, Y.; Hallou, A.; Charles, B. *J. Phys. Chem.* **2003**, *107*, 2011; Gauduel, Y.; Hallou, A. *Res. Chem. Intermed.* **2001**, *27*, 359; Gauduel, Y.; Sander, M.; Gelabert, H. *J. Phys. Chem. A* **1998**, *102*, 7795.

48. Sheu, W.-S.; Rossky, P. J. *Chem. Phys. Lett.* **1993**, *202*, 186 and 233; *J. Am. Chem. Soc.* **1993**, *115*, 7729; *J. Phys. Chem.* **1996**, *100*, 1295.

49. D. Borgis and A. Staib, Chem. Phys. Lett. **230**, 405 (1994); J. Chim. Phys. **93**, 1628 (1996); J. Phys.: Condens. Matter **8**, 9389 (1996); J. Mol. Struct. **436**, 537 (1997); J. Chem. Phys. **103**, 2642 (1995)

50. Bradforth, S. E.; Jungwirth, P. *J. Phys. Chem. A* **2002**, *106*, 1286.

51. Spezia, R.; Nicolas, C.; Archirel, P.; Boutine, A. *J. Chem. Phys.* **2004**, *120*, 5262.

**Figure captions.**

**Fig. 1.**

(a) Normalized 200 nm pump - 800 nm probe transient absorbance kinetics of electron generated by photoexcitation of (i) 0.74 M and (ii) 1.84 M (saturated) aqueous sodium sulfate. Note the logarithmic time scale. These kinetic traces were normalized at delay time $t=5$ ps; the difference in the short-term kinetics is due to the blue shift in the electron spectrum that increases with the ionic strength (sections 3.3 and 3.6.1). The difference in the long-term kinetics is due to ionic strength effect on the dynamics of geminate $\left(SO_4^-; e_{aq}^-\right)$ pairs (sections 3.6). (b) 200 nm pump - 800 nm probe kinetics obtained for aqueous solutions of 9.3 M sodium perchlorate *(empty circles, to the left)* and 5.43 M sodium chloride *(solid line, to the right)*. The vertical bars in this plot indicate 95% confidence limits.

**Fig. 2.**

(a) *Solid lines:* Absorption spectra of 0.7 mM NaI vs. sodium perchlorate concentration (1 cm optical path cell). The molar concentrations of the perchlorate *(from right to left)* are 0, 1, 2.5, 5, 7, and 9 M (the direction of the increase in the ionic strength $I$ is indicated with an arrow in the plot). Dashed lines are the Gaussian fits to these absorption spectra. (b) Negative shift $\Delta\lambda$ of the band maximum for aqueous iodide vs. the nominal ionic strength of sodium perchlorate *(circles)* and sodium sulfate *(squares)*. Empty symbols indicate the data obtained in the present work; filled symbols are data taken from ref. [30].

**Fig. 3.**

Molar absorptivities of CTTS anions (given in the units of $M^{-1}$ $cm^{-1}$) at the laser photoexcitation wavelength vs. the salinity of photolysate (that was changed by addition of NaClO$_4$ and Na$_2$SO$_4$). The filled symbols are spectrophotometric data (e.g., Fig. 2(a)); the empty symbols are laser transmission data (sections 2 and 3.4). (a) Molar absorptivity of aqueous iodide at 248 nm. The circles are for sodium perchlorate and the squares are for sodium sulfate solutions. The solid line is an exponential fit. (b) Molar absorptivities of aqueous hydroxide at 193 nm vs. [NaClO$_4$] *(to the left and to the bottom)* and of sulfite at 248 nm vs. [Na$_2$SO$_4$] *(to the right and to the top)*. The solid lines are linear fits.

**Fig. 4.**

Prompt visible transient absorption spectra of solvated electron obtained in radiolysis of N$_2$-saturated (a) sodium sulfate and (b) sodium perchlorate solutions (section 3.3). The molar concentrations of these salts are given in the plot. Solid lines are Lorentzian-Gaussian fits of constant width (see section 3.3).

**Fig. 5.**

Ionic strength dependence of (a) solvated electron absorbance at the maximum (estimated from the data of Fig. 4), (b) position of the band maximum of the solvated electron in energy units, and (c) electron absorbance at 700 nm. The data for sodium perchlorate and sodium sulfate are indicated with open circles and squares, respectively.

**Fig. 6.**

Transient absorption of photoelectrons (observed at 700 nm) vs. absorbed laser power. The direction of increase in the nominal ionic strength $I$ of photolyzed solution is indicated by an arrow. The solid lines drawn through the data points are linear fits. (a) Electron absorbance in 248 nm laser photolysis of 38.5 mM sulfite in *(from top to bottom)* 0.36, 0.73, 1.19 and 1.73 M sodium sulfate. (b) Electron absorbance in 248 nm laser photolysis of 2 mM iodide in *(from top to bottom)* 0.174 M *(squares),* 0.353 M *(triangles),* 0.59 M *(diamonds),* 0.9 M (circles) and 1.7 M *(upturned triangles)* sodium sulfate. (c) Electron absorbance in 193 nm laser photolysis of hydroxide in *(from top to bottom)* 2.1, 4.5, and 9 M sodium perchlorate vs. the laser power $I_{abs}(OH^-)$ absorbed by the hydroxide. [KOH] was 0.75 mM *(squares)* and 1.5 mM *(triangles)* for 2.1 M NaClO$_4$ series, 0.75 mM *(squares)* and 1.5 mM *(upturned triangles)* for 2.1 M NaClO$_4$ series, 2 mM (circles) for 4.5 M NaClO$_4$ series, and 1 mM (triangles) and 2 mM *(diamonds)* for 9 M NaClO$_4$ series.

**Fig. 7.**

(a) Ionic strength dependence of relative molar absorptivity $\varepsilon_{700}/\varepsilon_{700}(I=0)$ of solvated electron at 700 nm in sodium perchlorate *(circles)* and sodium sulfate *(squares)* solutions obtained as explained in section 3.5. The solid and dashed lines (for perchlorate and sulfate, respectively) were calculated from the data of Fig. 5(b) assuming the constancy of the oscillator strength of the corresponding optical transition. (b) Relative quantum yield of electron photodetachment vs. nominal ionic strength for 248 nm photoexcitation of *iodide (filled circles)* and sulfite *(empty diamonds)* in sodium sulfate solutions and iodide *(empty triangles)* in sodium perchlorate solutions and 193 nm photoexcitation of perchlorate *(empty circles)* and hydroxide *(empty squares)* in sodium perchlorate solutions. The linear plots *(solid lines marked (i), (ii), and (iii))* correspond to decreases of 11%, 9.8%, and 6.2% per 1 M ionic strength, respectively.

**Fig. 8.**

(a) 200 nm pump - 800 nm probe transient absorption kinetics of electron in photolysis of 30 mM NaBr in solution containing 2.4 to 9.25 M sodium perchlorate. The molarity of NaClO$_4$ is indicated in the color scale. Absorption signal (i) is from 9.3 M perchlorate with no bromide added (under the same excitation conditions). The absorption of 200 nm photoexcitation light by perchlorate in the bromide solutions was negligible. The vertical bars are 95% confidence limits. (b) The same kinetic data *(circles)* normalized at $t=5$ ps (the upper trace is for aqueous bromide with no NaClO$_4$ added). The same color coding used as in (a); the direction of the increase in the nominal ionic strength $I$ is indicated by an arrow. The solid lines drawn through the symbols are simulations using Shushin's theory (section 3.6.3) with the parameters given in Fig. 9.

**Fig. 9.**

Simulation parameters used to fit the data shown in Figs. 8(b) and 8S vs. the nominal ionic strength of the aqueous solution. (a) The escape probability of the photoelectron in 200 nm photoexcitation of bromide in sodium sulfate *(squares)* and sodium perchlorate

*(circles)* solutions. **(b)** Dissociation *($W_d$, circles)* and recombination *($W_r$, squares)* rate constants for caged $(Br, e_{aq}^-)$ pairs in concentrated sodium perchlorate solutions.

**Fig. 10.**

Normalized 200 nm pump - 800 nm probe kinetics of photoelectron generated by photolysis of **(a)** 40 mM $Na_2SO_3$ and **(b)** 200 mM KOH in concentrated sodium sulfate solutions. The molarity of sulfate is given in the color scale. These kinetics were normalized at *t*=5 ps. See Fig. 9S for non-normalized kinetics.

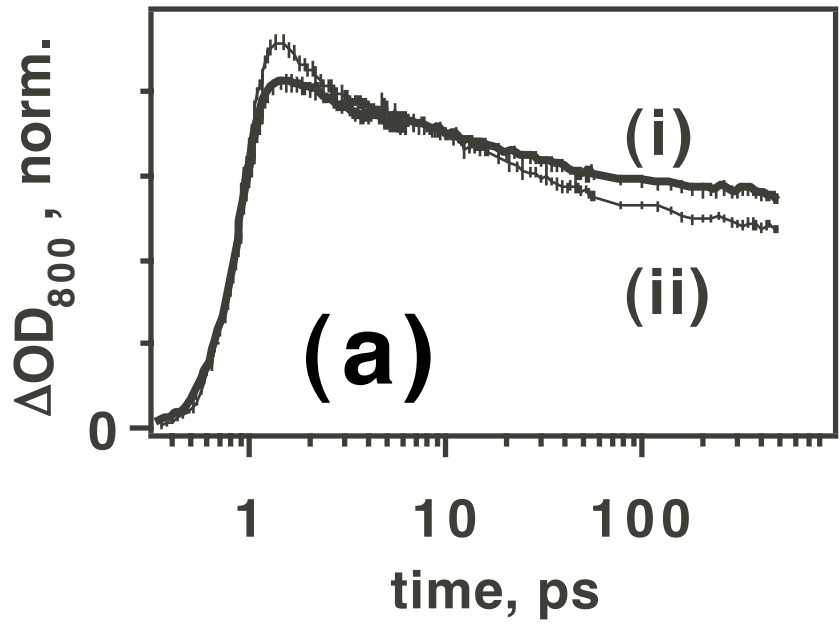

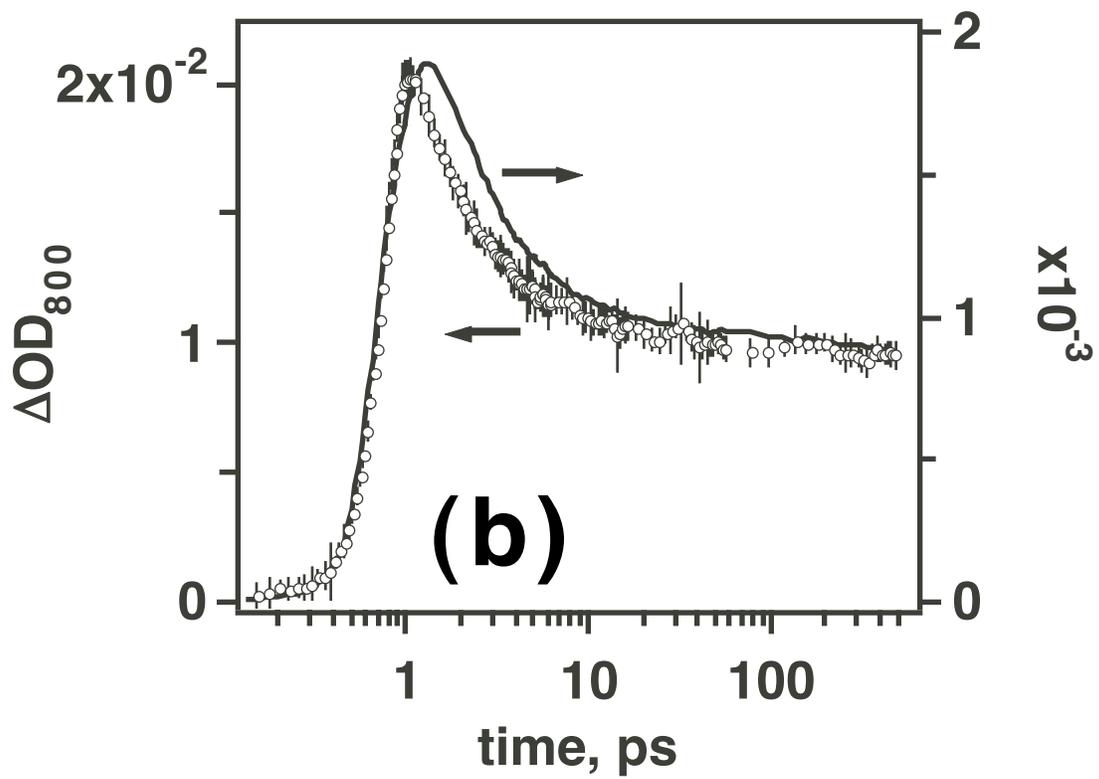

Fig. 1; Sauer et al.

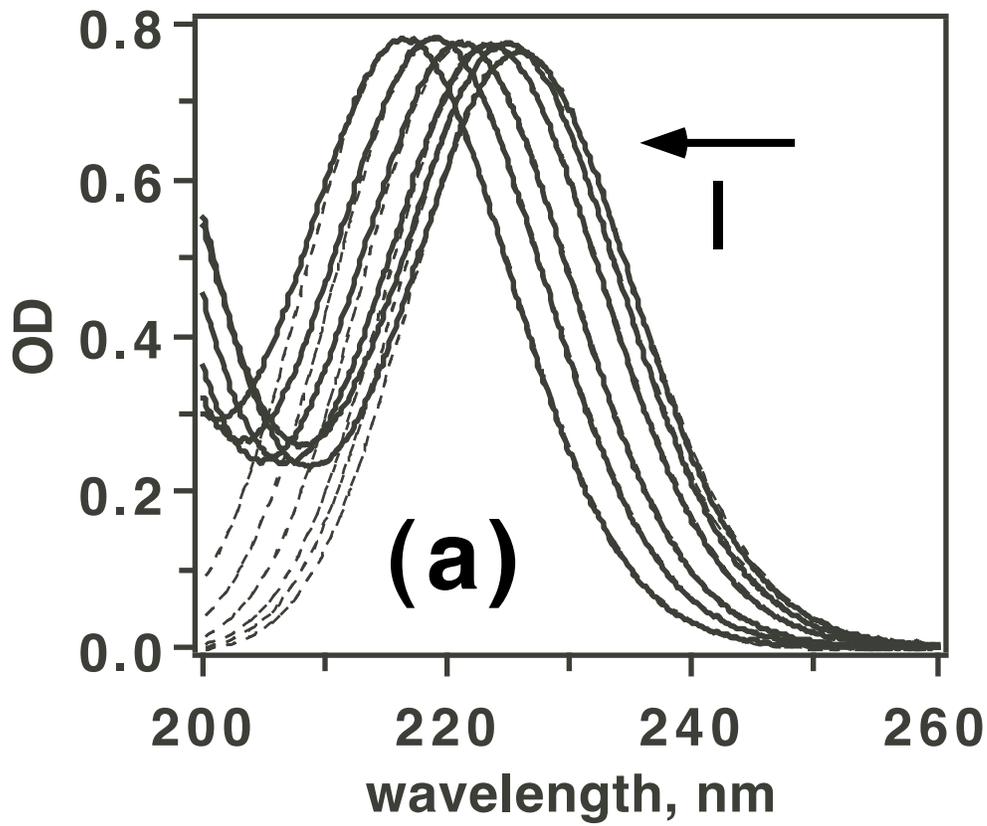
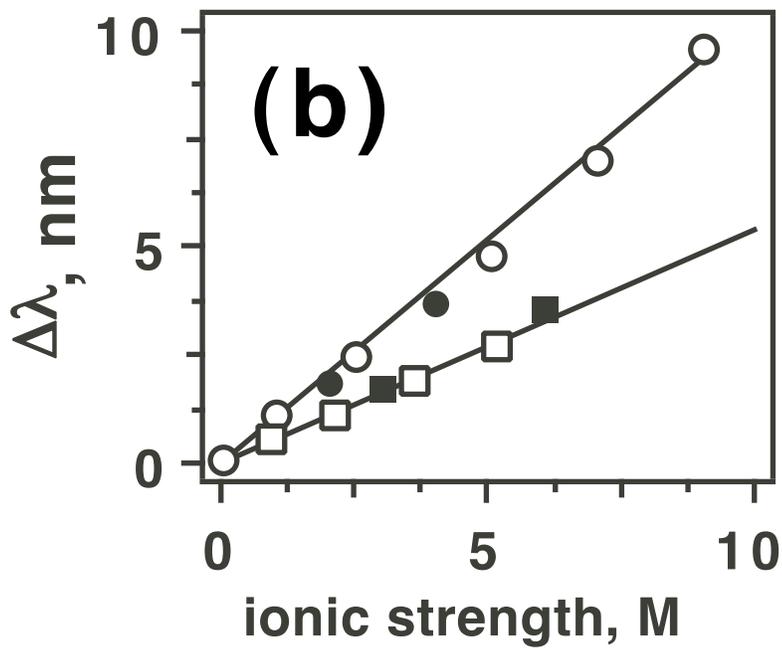

Fig. 2; Sauer et al.

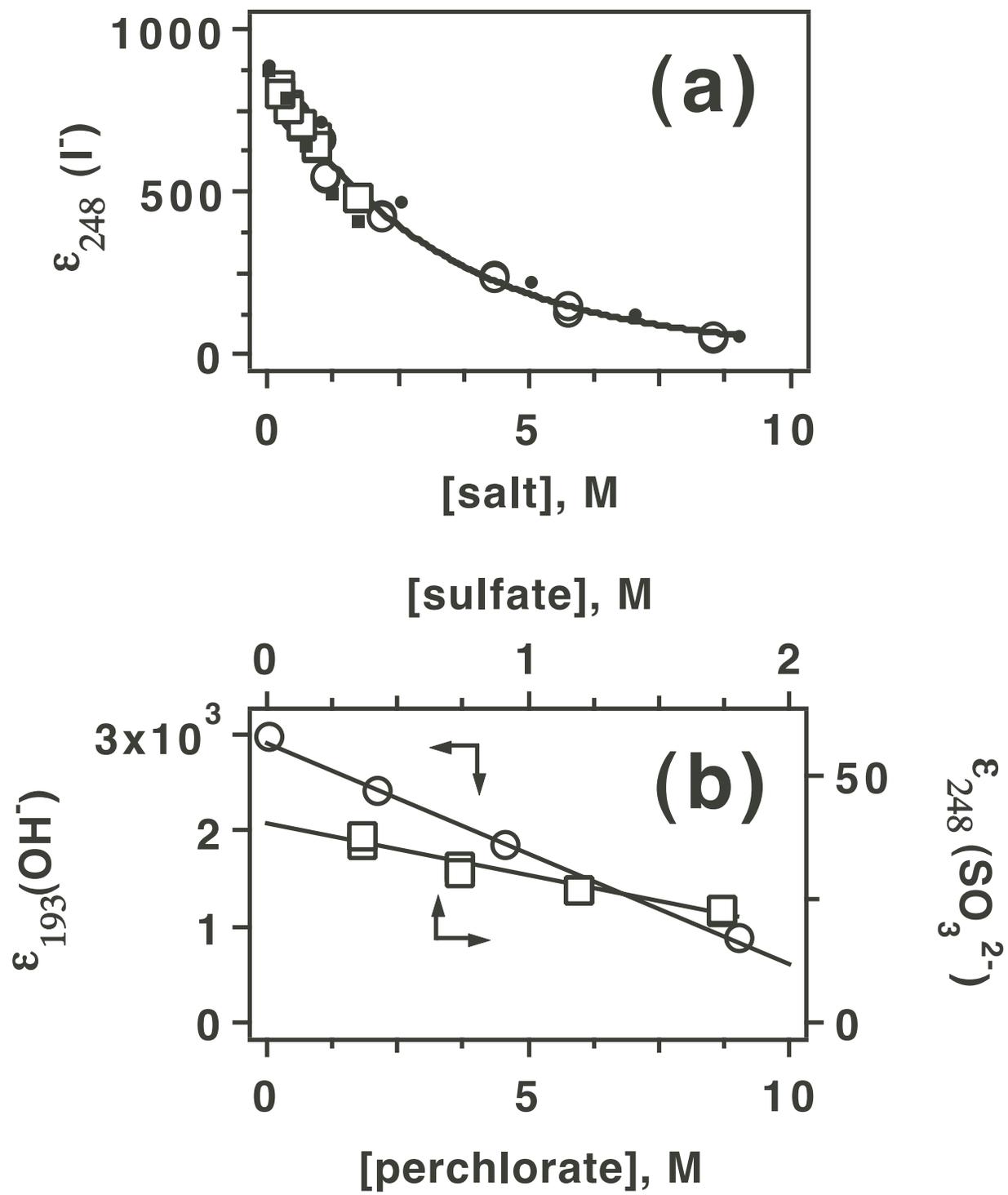

Fig. 3; Sauer et al.

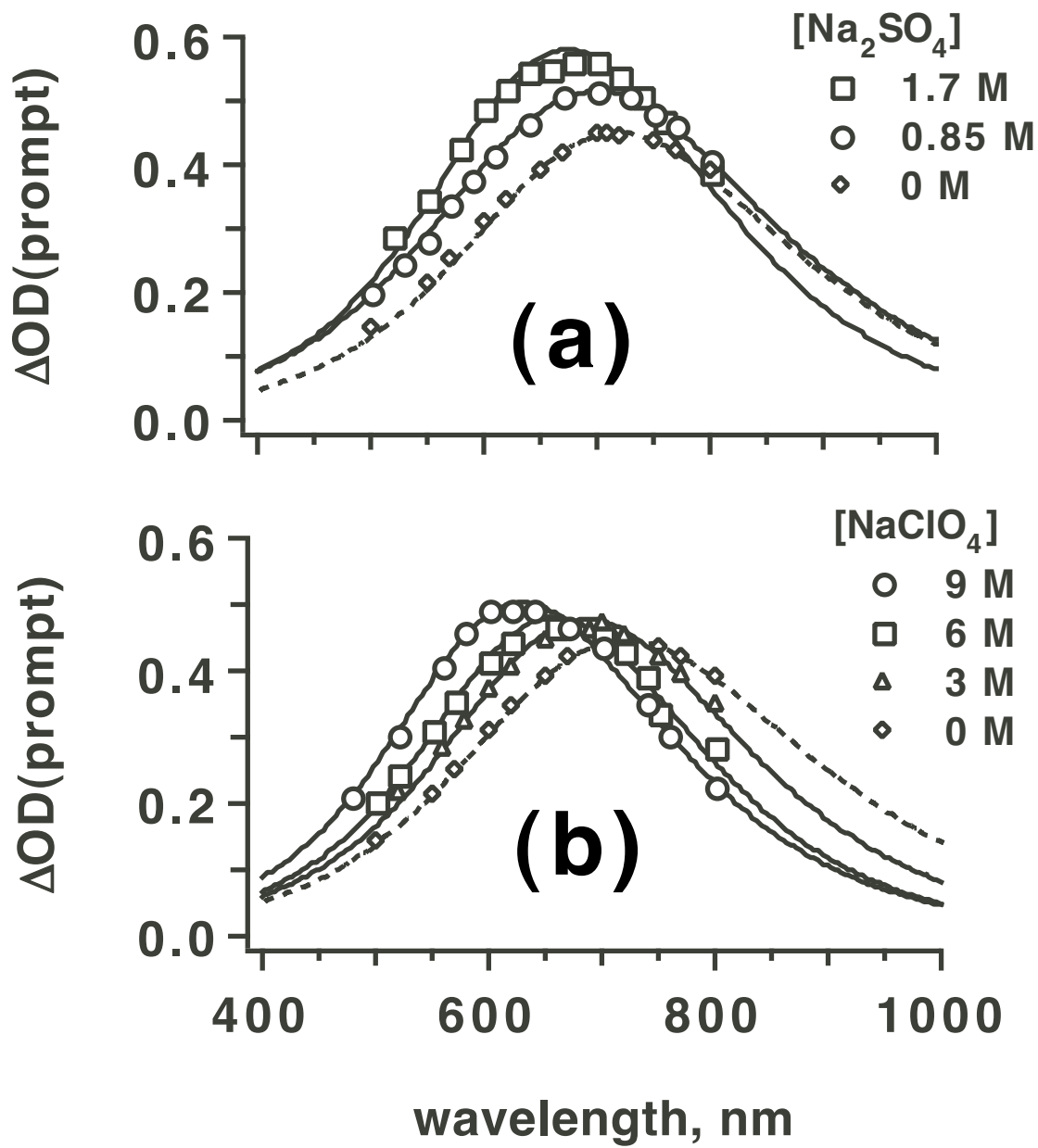

Fig. 4; Sauer et al.

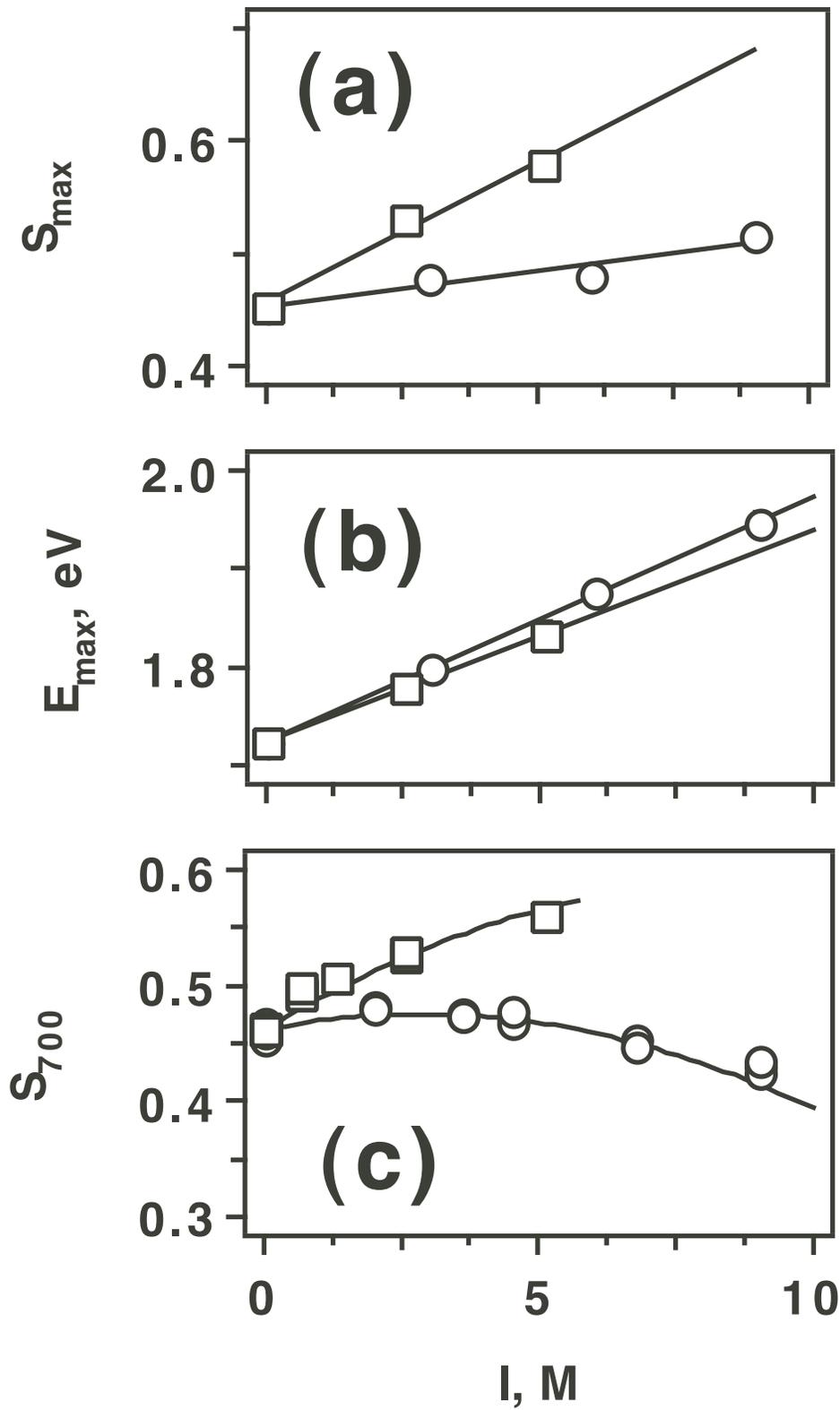

Fig. 5; Sauer et al.

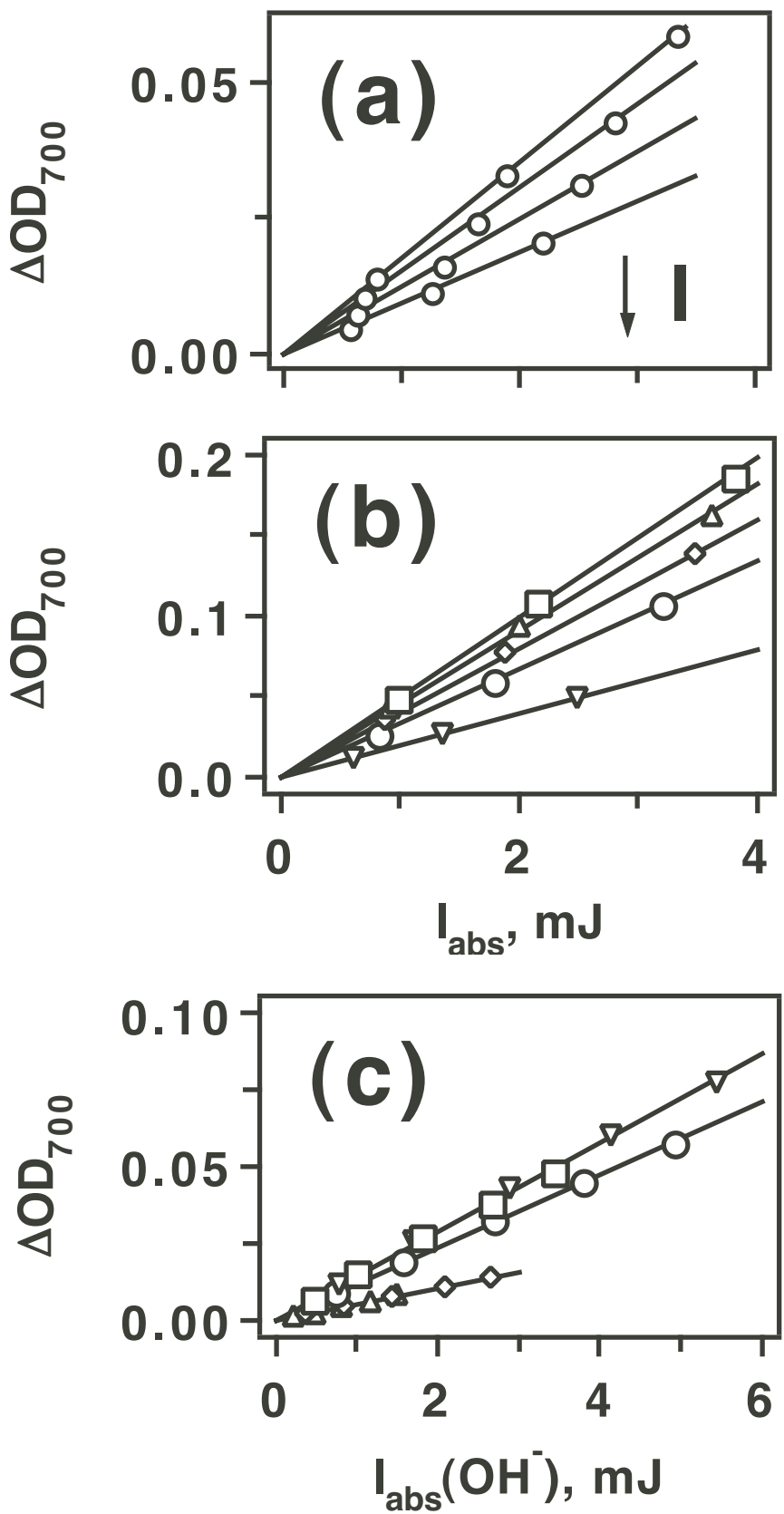

Fig. 6; Sauer et al.

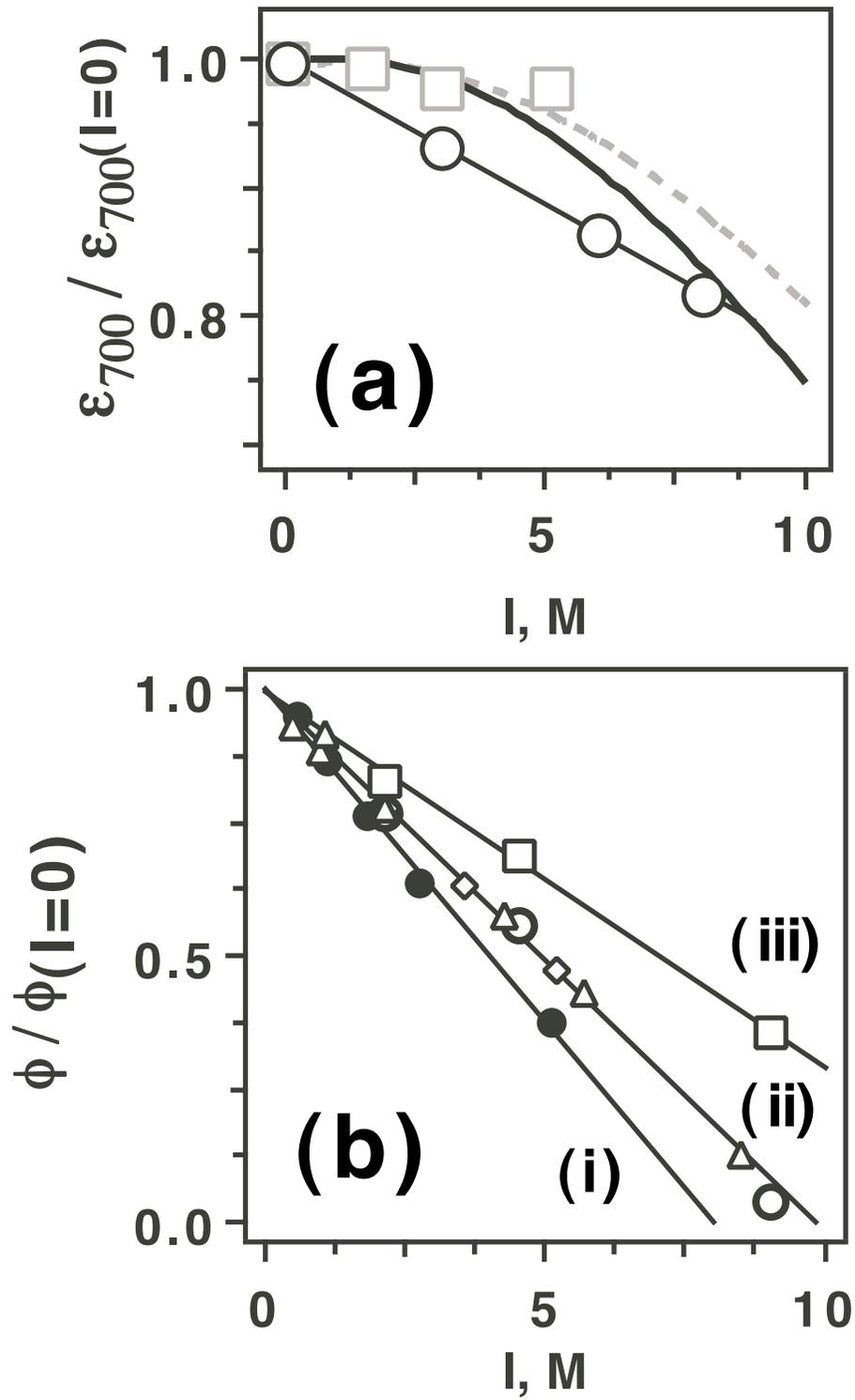

Fig. 7; Sauer et al.

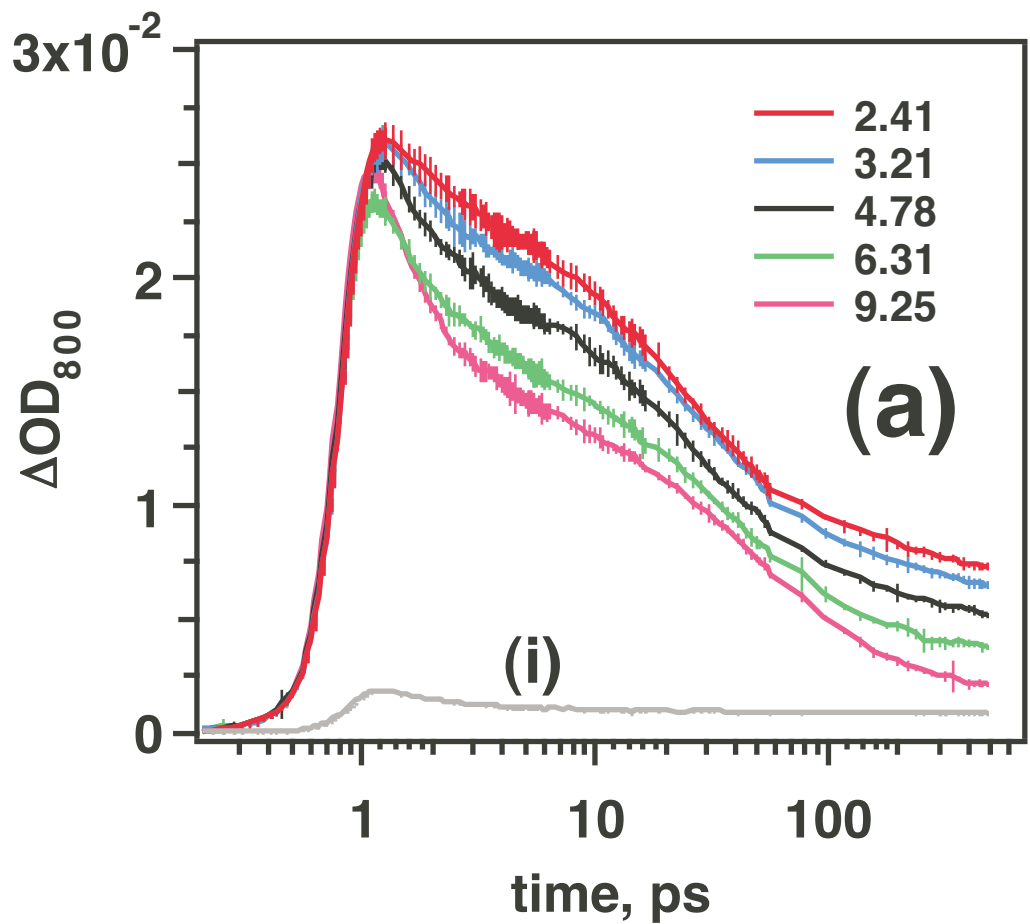
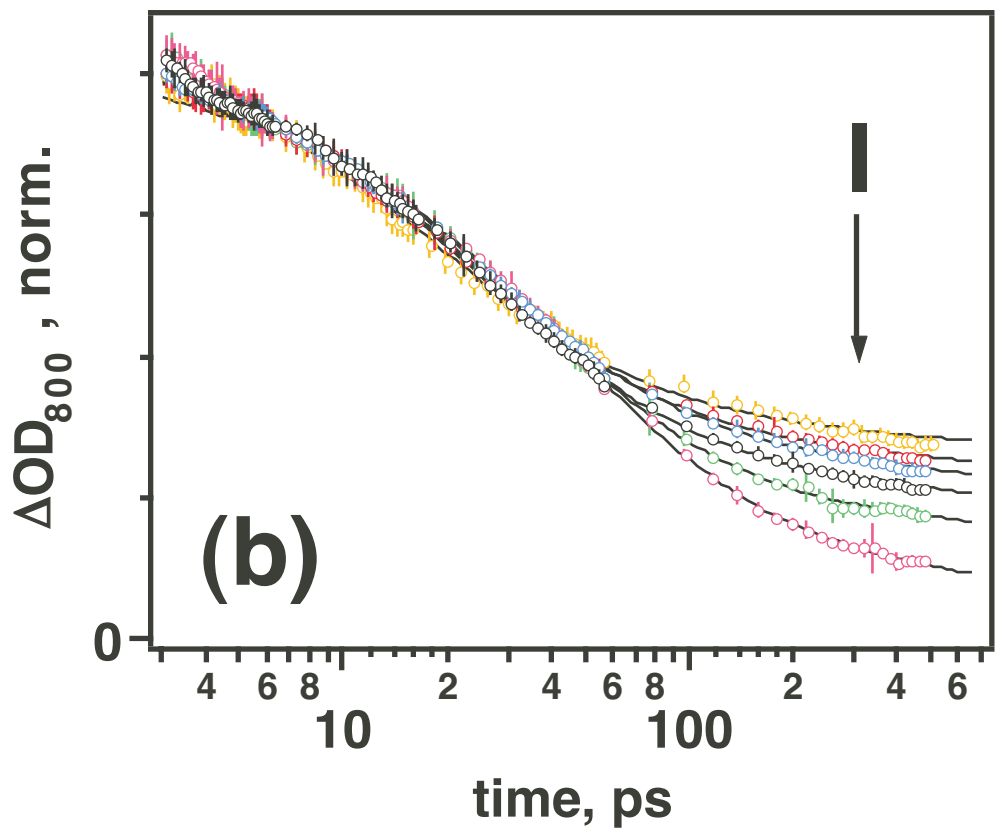

Fig. 8; Sauer et al.

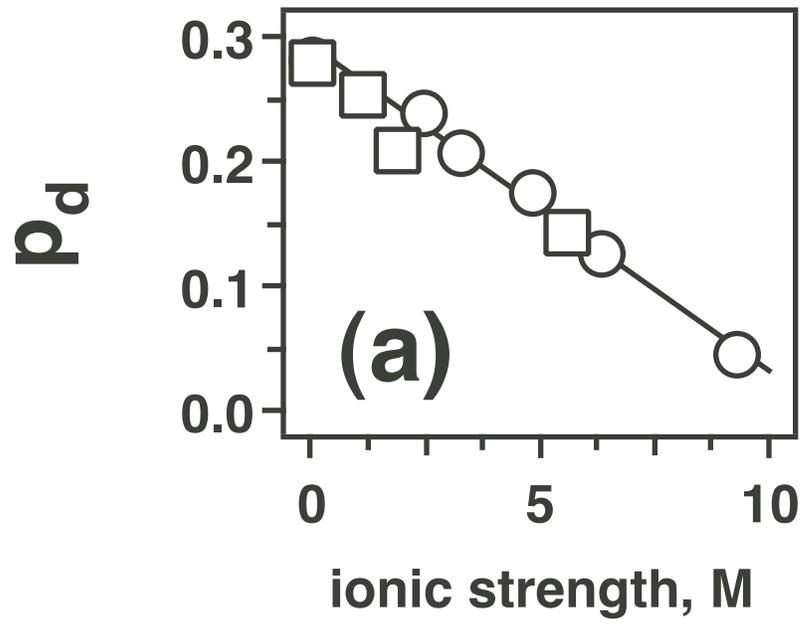
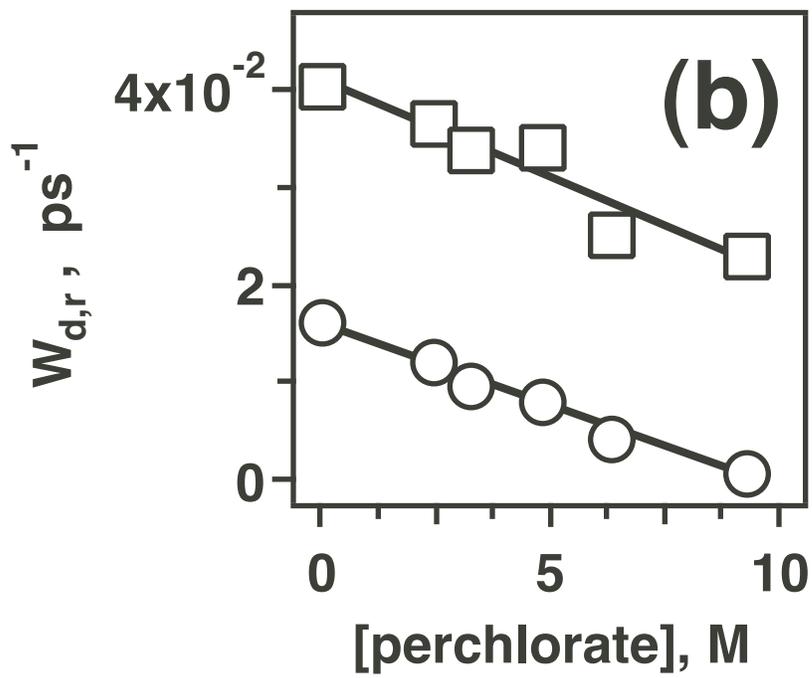

Fig. 9; Sauer et al.

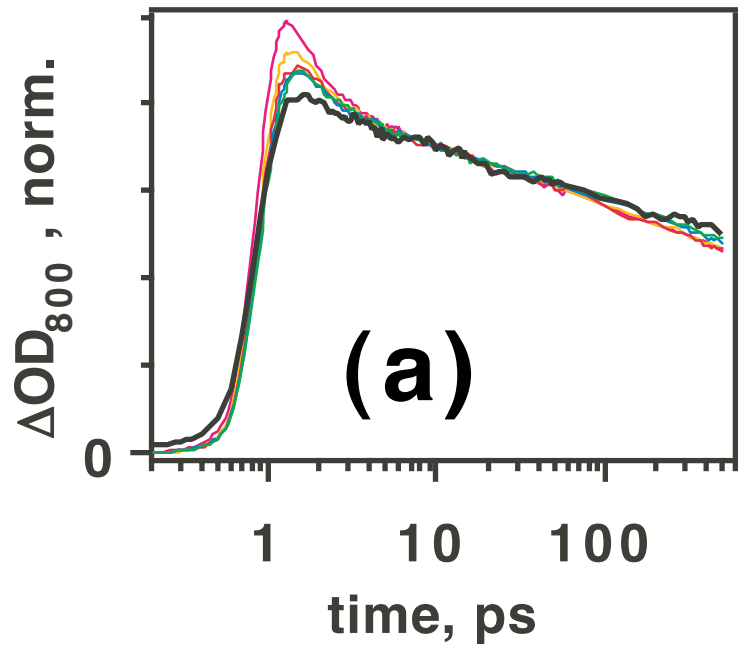

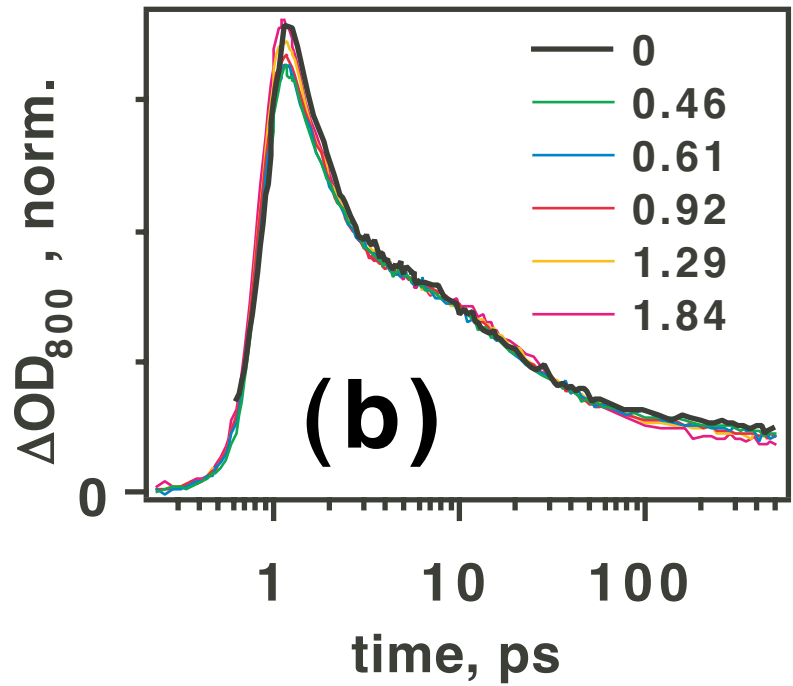

Fig. 10; Sauer et al.

**SUPPLEMENTARY MATERIAL**           **jp0000000**

*Journal of Physical Chemistry A, Received * 2004*

**Supporting Information.**

**Captions to figures 1S to 9S.**

**Fig. 1S.**

(a) Viscosity *(to the left)*, density *(to the right)* and (b) ion diffusivity vs. molarity of aqueous sodium sulfate (from the data of refs. [17] and [21]).

**Fig. 2S.**

(a) Mean ionic coefficient *(empty squares)* [17] and (b) the fraction of solvent-separated $\left(Na^+ \cdot OH_2 \cdot SO_4^-\right)$ ion pairs *(to the left)* and nominal ionic strength $I$ of the solution *(to the right)* vs. molarity of aqueous sodium sulfate (data of ref. [23] were used to obtain these curves).

**Fig. 3S.**

Same as Fig. 1S, for sodium perchlorate. Reference data [17,21] were used to generate these plots.

**Fig. 4S.**

(a) Power dependence of photoelectron optical absorbance (at 800 nm) observed 10 ps after short-pulse 200 nm laser excitation of 0.74 M aqueous sodium sulfate. This linear plot indicates that the excitation of sulfate anion is single-photon (note that water ionization by the 200 nm light is biphotonic). An estimate of 0.406 for the prompt yield of thermalized electron in 200 nm photoexcitation of $SO_4^{2-}$ is calculated from these data. (b) A demonstration of the invariance of the kinetic traces for 0.74 M sulfate with the incident laser power: the kinetic traces for 0.35 µJ *(circles)* and 1.63 µJ *(squares)* photoexcitation are identical within the experimental error (the vertical bars in the plot indicate 95% confidence limits for the data points).

**Fig. 5S.**

(a) Concentration dependence of 193 nm absorption $OD_{193}$ of the photolytic cell for hydroxide in $NaClO_4$ solutions ($L$ is the optical path of the cell, which is 1.36 mm), as determined from the transmission of ArF laser light. The molar concentrations of $NaClO_4$ are given in the plot; the hydroxide molarity is given in the caption to Fig. 6(c). Non-zero offset of the linear plots is due to the absorbance of 193 nm light by perchlorate. (b) Relative quantum yield $\Phi_{rel}$ of the photoelectron as a function of the nominal ionic strength $I$. Photoexcitation wavelengths, CTTS anions, and salts used to change the ionic strength of the photolysate are specified in the legend given to the right of the plot.

**Fig. 6S.**

Transient optical absorbance (700 nm) in 248 nm laser photolysis of 2 mM NaI in (a) 0 M, (b) 3 M, and (c) 6 M NaClO$_4$. Bold line *(red)* kinetic traces are from N$_2$-saturated solutions *(to the right)*; thin line *(green)* traces are from CO$_2$-saturated solutions *(to the left)*; these two sets were obtained under the identical photoexcitation conditions. Stronger absorption signals correspond to higher fluence of the incident 248 nm photons *(from bottom up):* 0.022, 0.04, and 0.06 J/cm$^2$. CO$_2$ was used as an electron scavenger; the progress of this scavenging can be observed in the 590 nm absorbance traces (dashed *(orange)* lines plotted in (b) and (c); *to the left*). I$_2^-$ does not absorb the 590 nm analyzing light. At 700 nm, in CO$_2$-saurated solutions one can observe both the rapid decay of the solvated electron and the slow formation of I$_2^-$ by rxn. (3).

**Fig. 7S.**

200 nm pump - 800 nm probe transient absorption kinetics for photoelectron in aqueous solutions containing sodium bromide (which is excited by this 200 nm light) and sodium sulfate (used to vary the ionic strength of the photolysate only). For convenience, these kinetic traces were normalized at $t=5$ ps. Molar concentrations of Na$_2$SO$_4$ and millimolar concentrations of NaBr are indicated above the plot. The absorption of 200 nm photons by sulfate was negligible in the presence of > 20 mM bromide. Rxn. (3) is too slow to change the dynamics of $\left(Br, e_{aq}^-\right)$ pairs in < 1 ns, even for 0.16 M Br$^-$ solution. Note the invariance of the normalized kinetics in 1.84 M (saturated) Na$_2$SO$_4$ solution with [Br$^-$]. The systematic increase in the short-lived thermalization "spike" is due to the blue shift in the absorption spectrum of thermalized electron with the ionic strength (sections 3.3 and 3.6.1).

**Fig. 8S.**

Replotted data of Fig. 7S for bromide in *(from top to bottom)* 0, 037, 0.62, and 1.84 M Na$_2$SO$_4$ solution *(circles)*. Solid lines are given by the global least squares fit of these kinetics obtained using Shushin's theory equations (section 3.6.3) with $W^{-1}$=19.5 ps, $\alpha$=0.552, and the escape probability $p_d$ given in Fig. 9(a).

**Fig. 9S.**

Same as Fig. 10; non-normalized optical densities are shown. Trace (i) is from 1.83 M sodium sulfate solution with no sulfite added; the photoexcitation of sulfate in the 40 mM sulfite solutions is negligible.

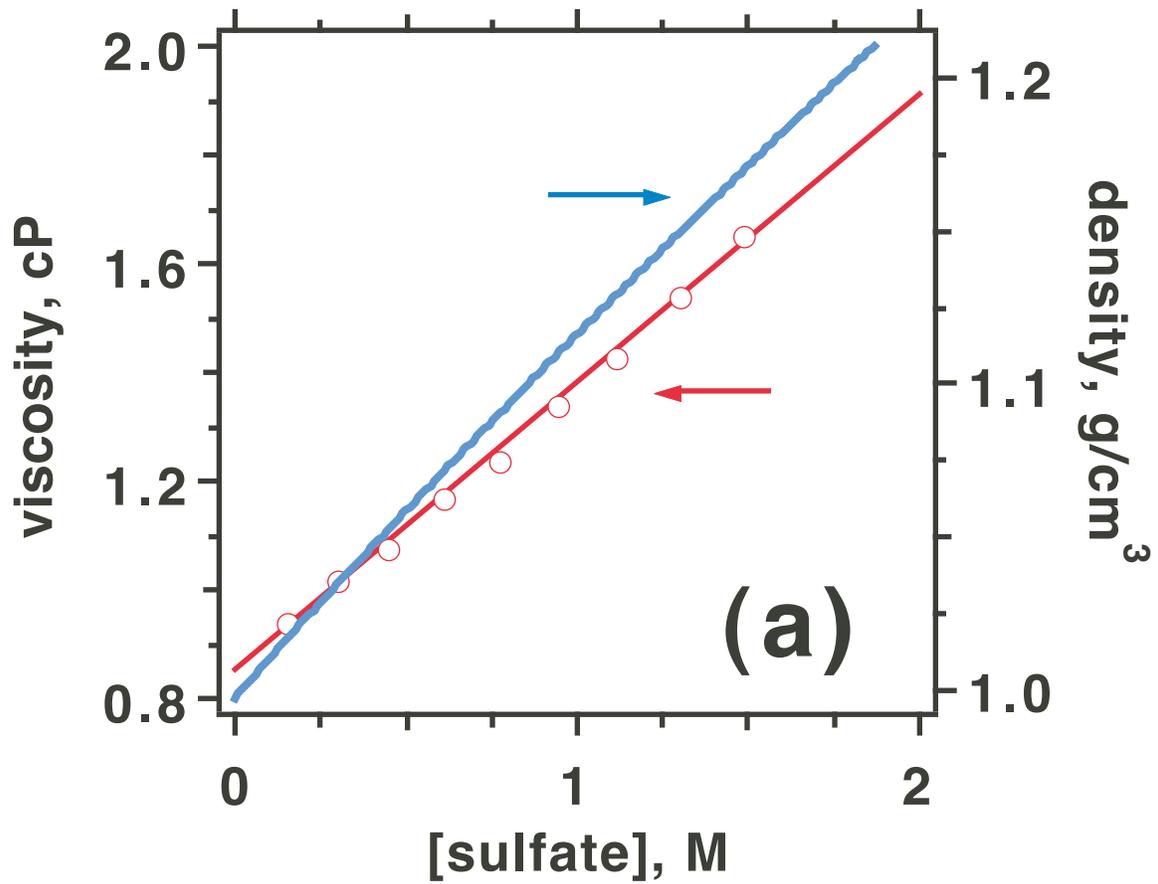

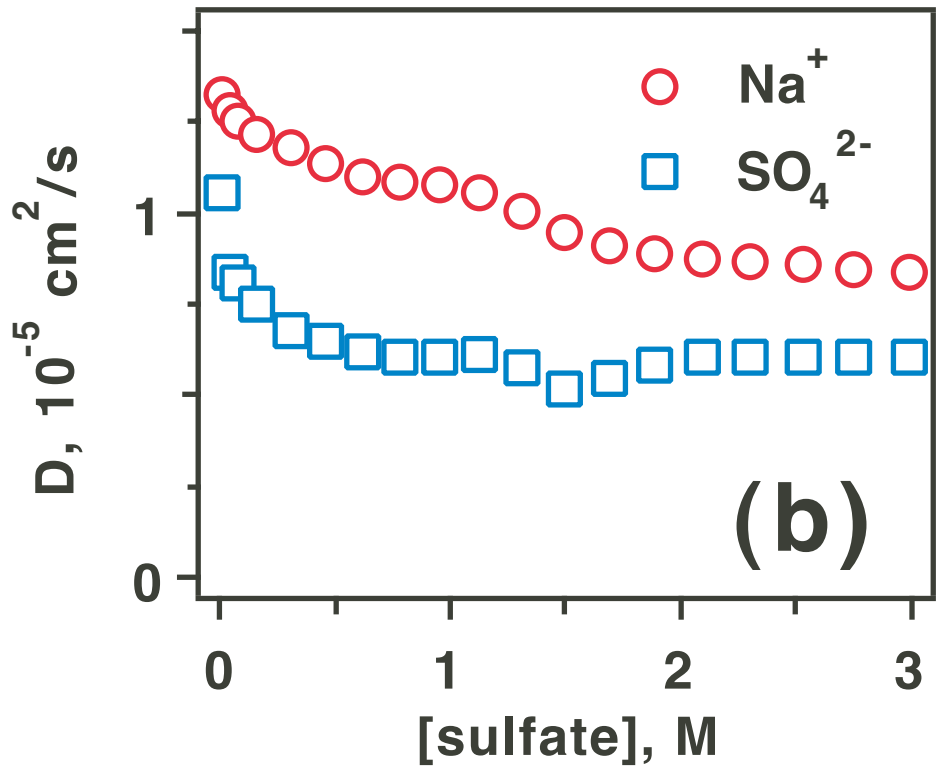

Fig. 1S; Sauer et al.

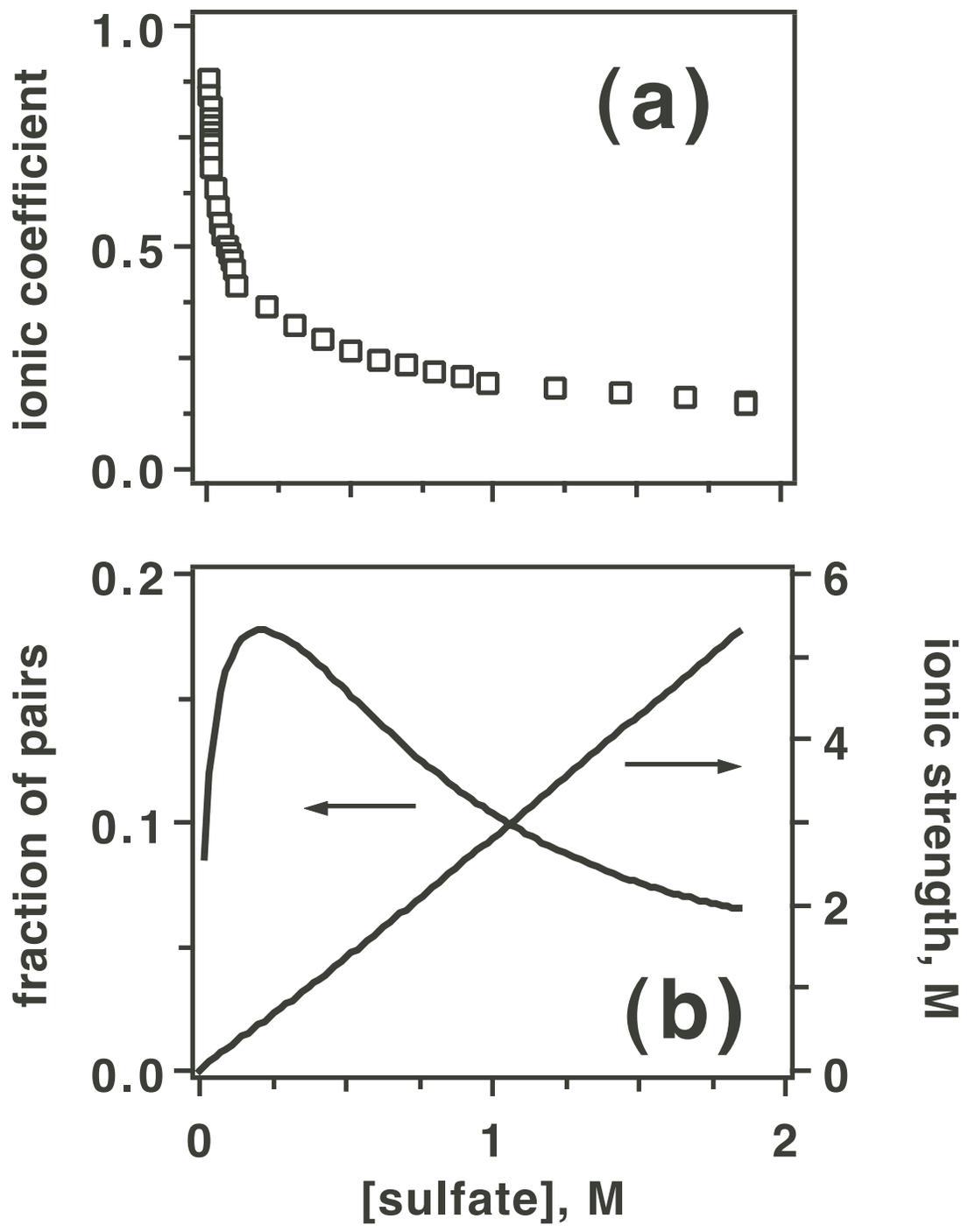

Fig. 2S; Sauer et al.

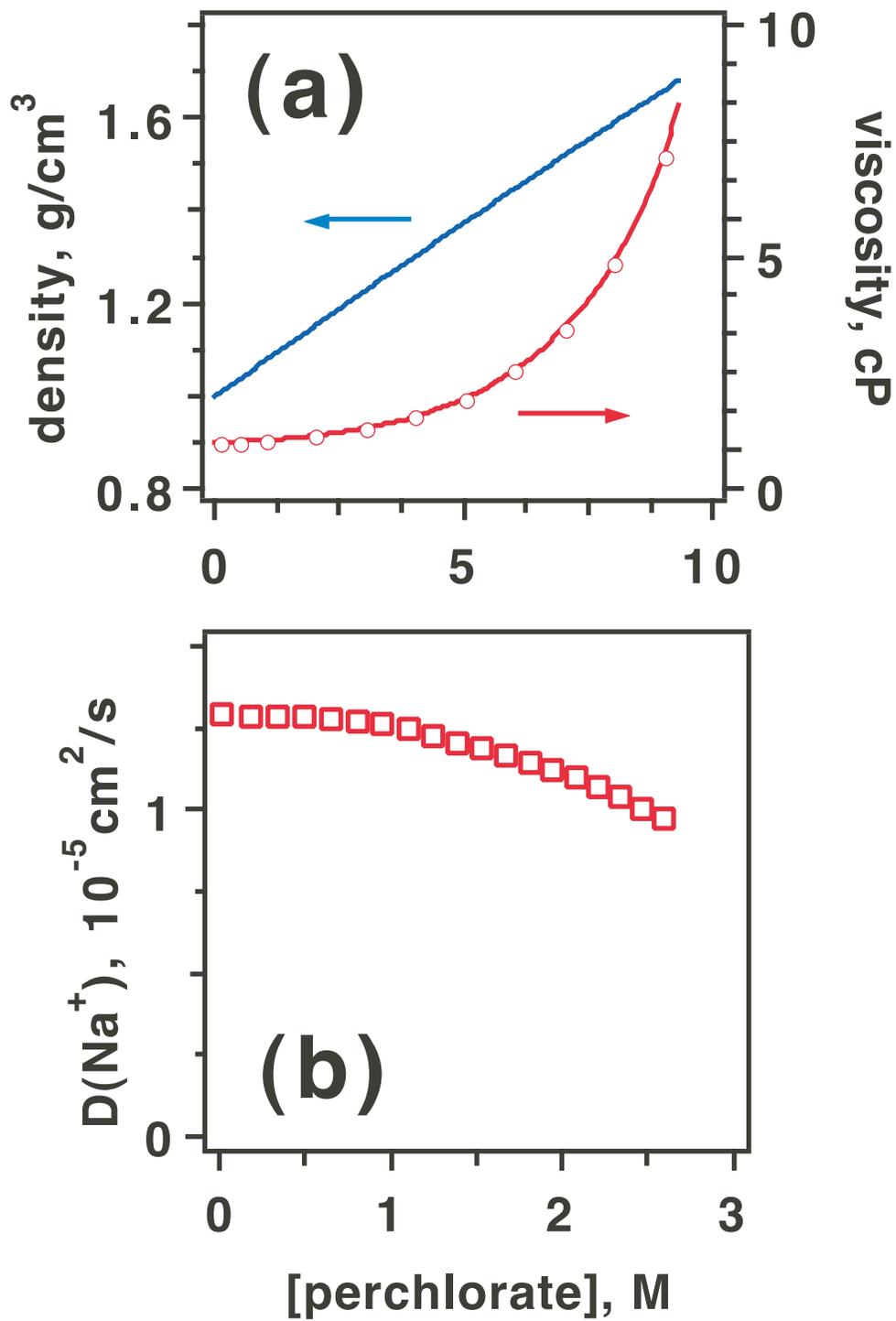

Fig. 3S; Sauer et al.

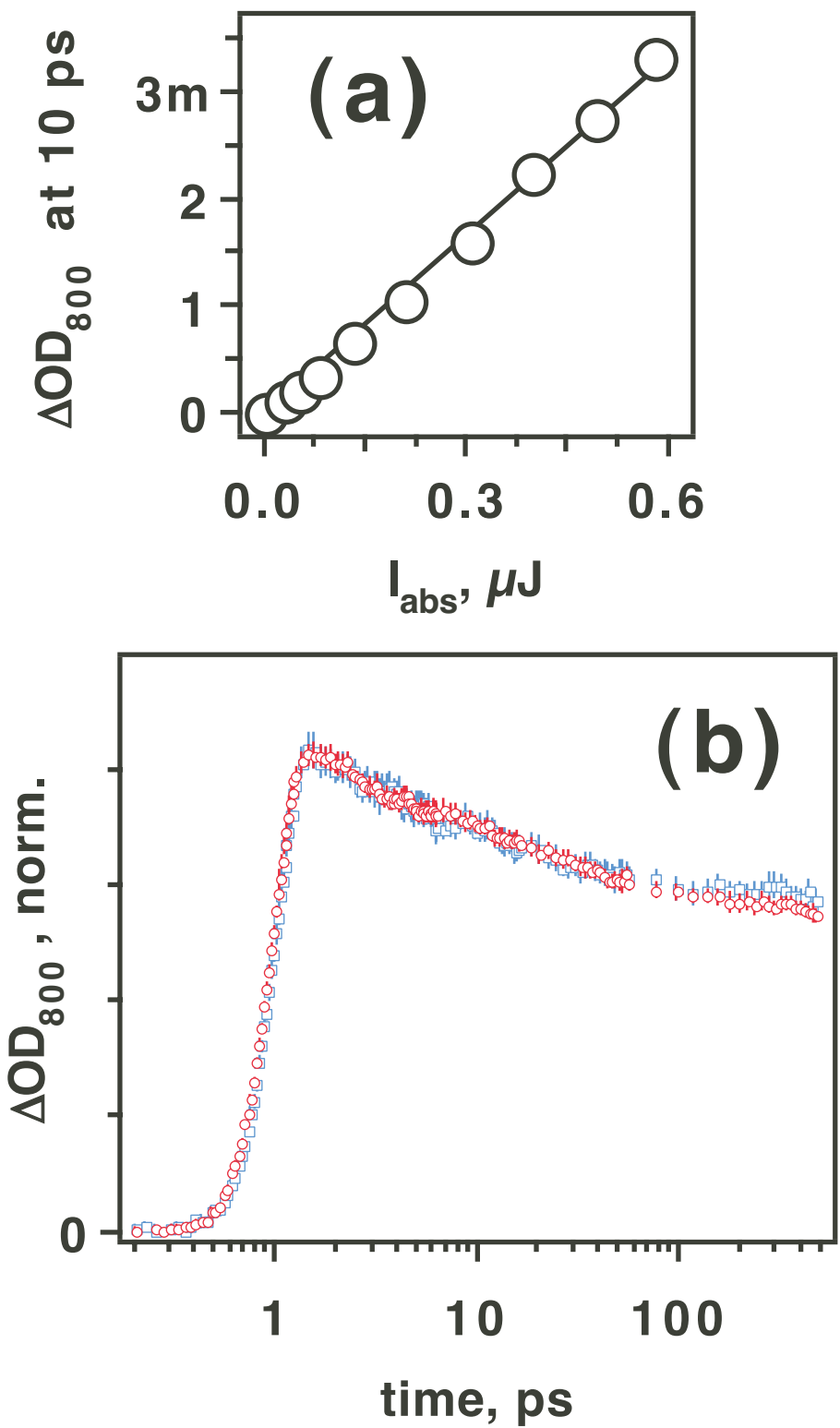

Fig. 4S; Sauer et al.

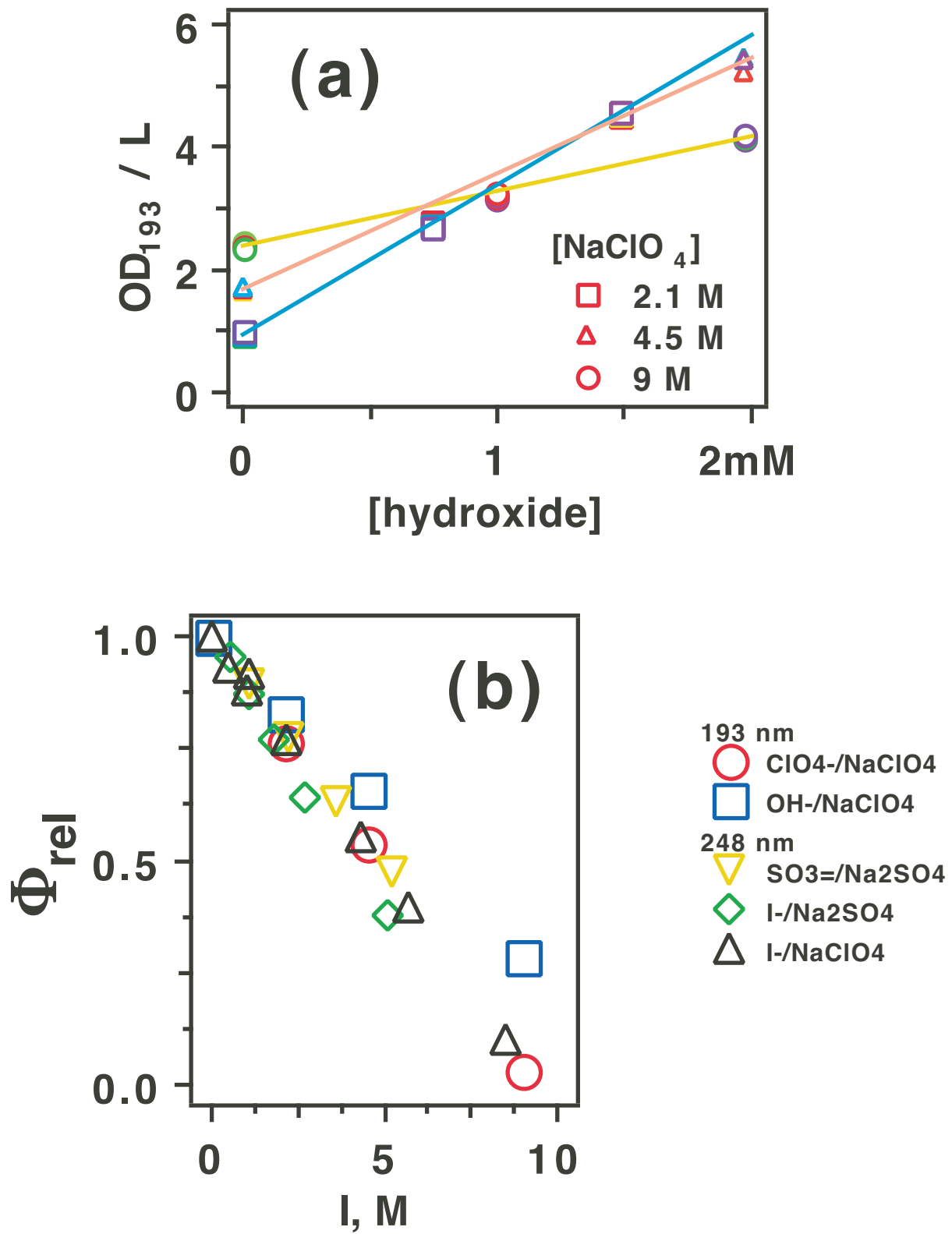

Fig. 5S; Sauer et al.

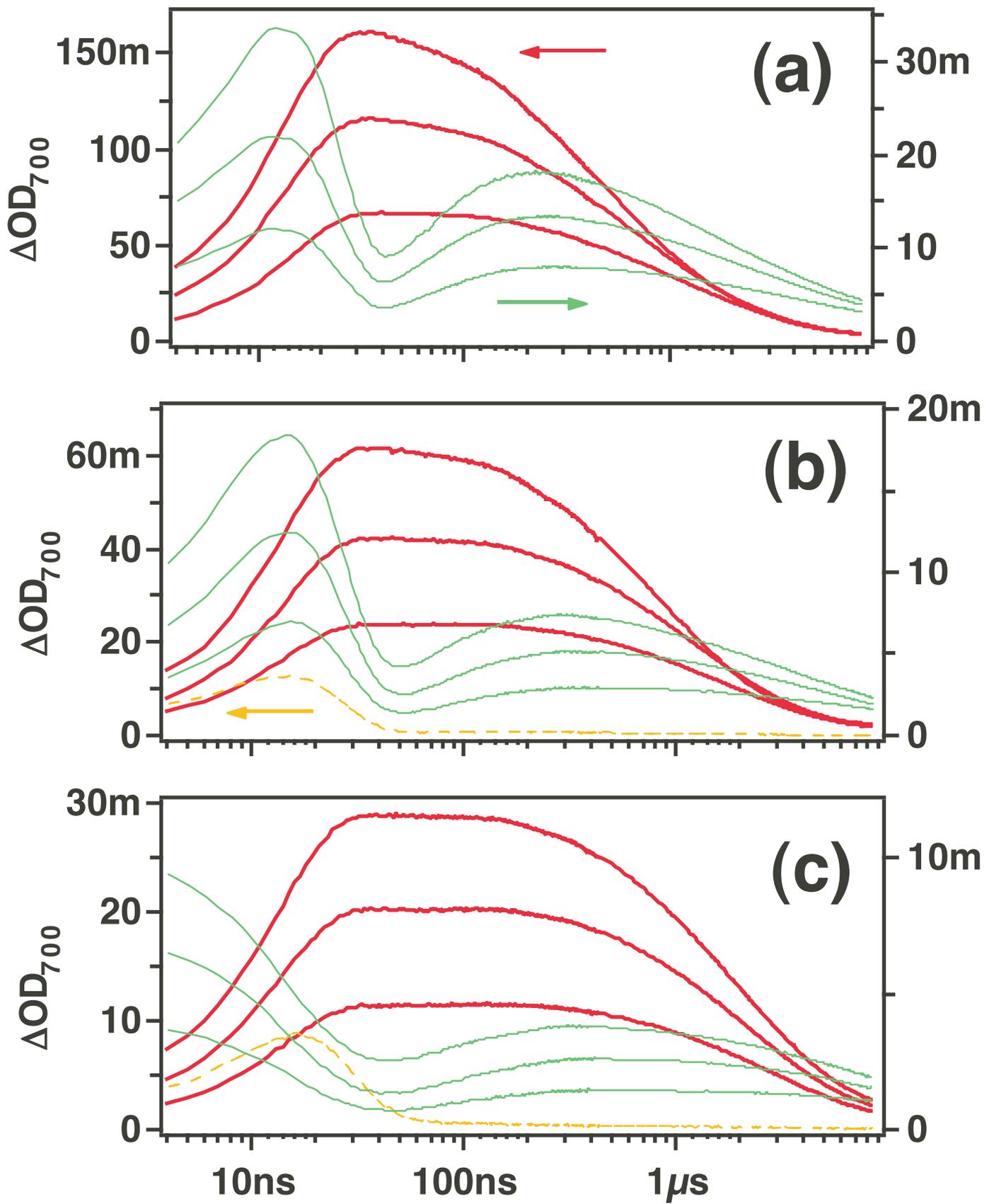

Fig. 6S; Sauer et al.

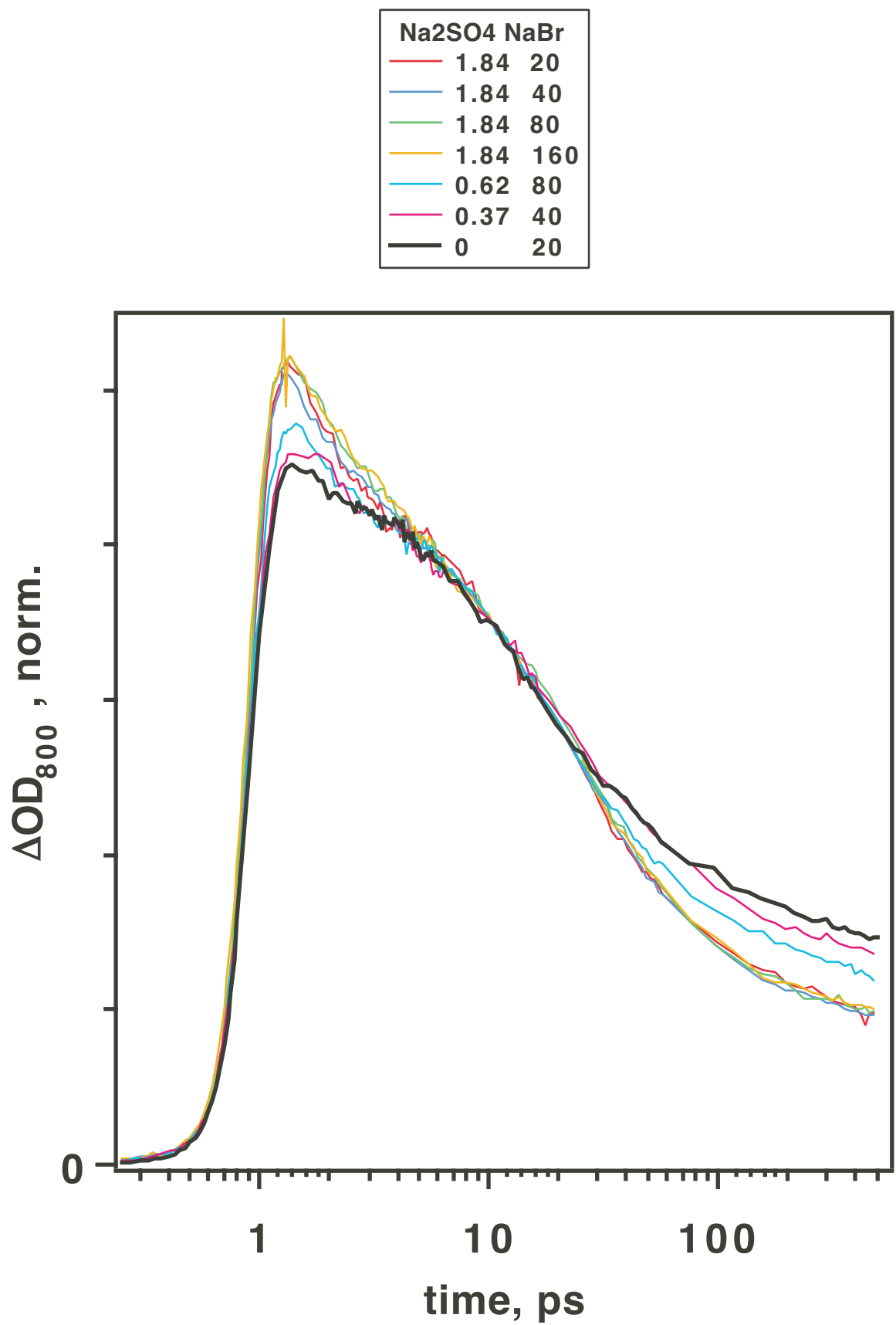

Fig. 7S; Sauer et al.

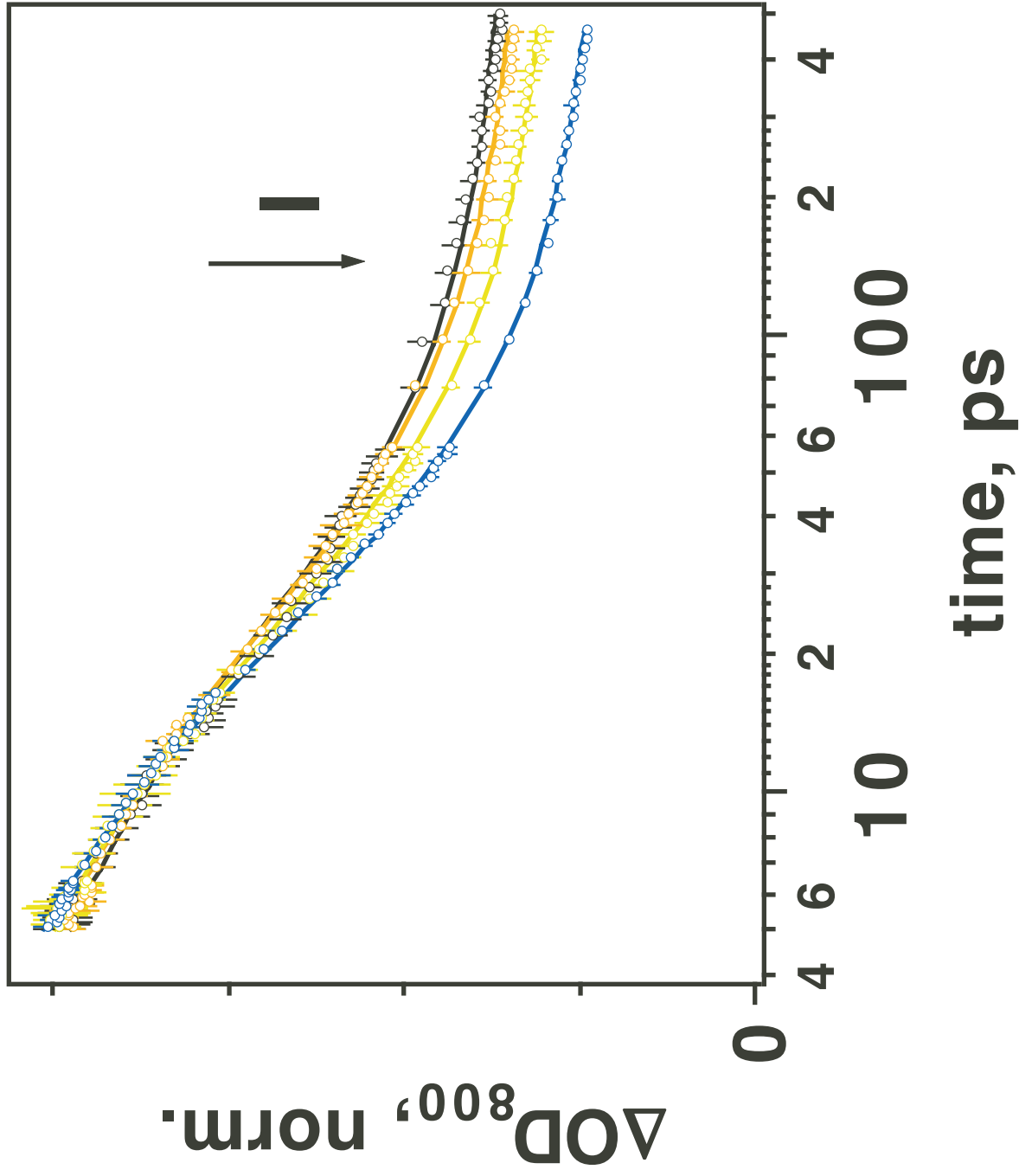

Fig. 8S; Sauer et al.

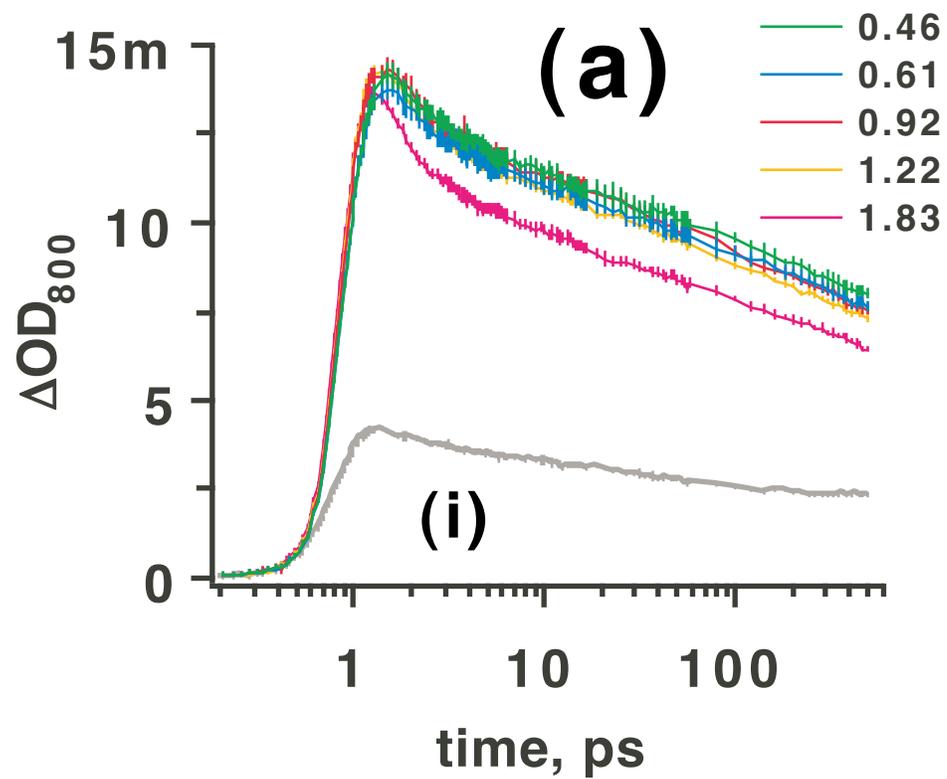
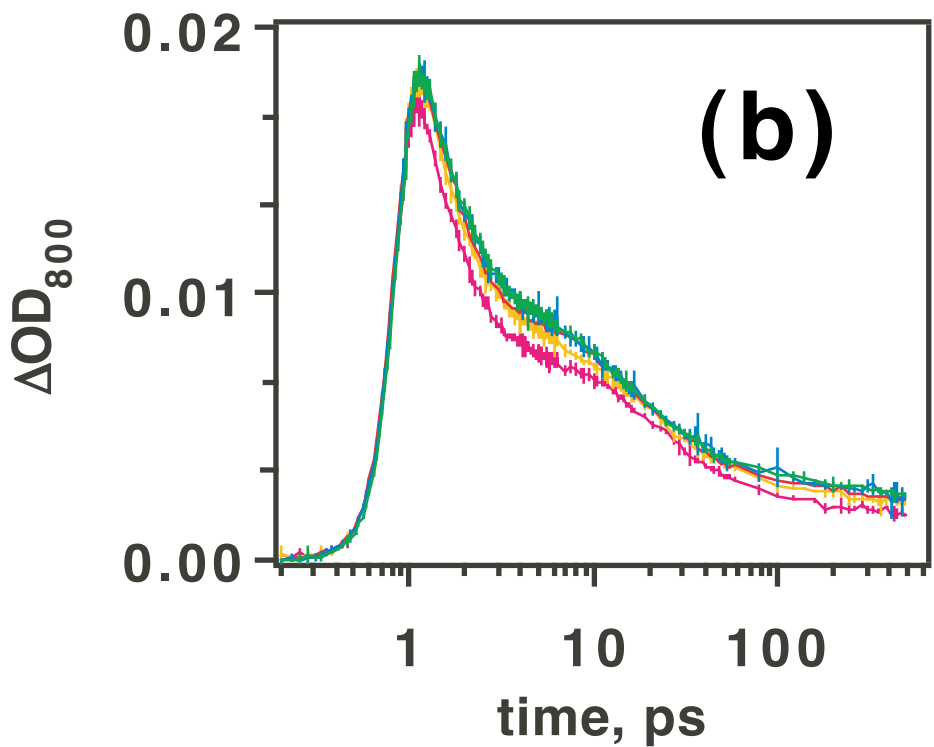

Fig. 9S; Sauer et al.